\documentclass[structabstract]{aa}  
\pdfoutput=1
\usepackage{graphicx}
\usepackage{amsmath}
\usepackage{txfonts}
\usepackage{natbib}
\usepackage{multirow}
\usepackage{longtable}
\usepackage{lscape}
\usepackage[draft]{hyperref}
\def\spitzer{{\em Spitzer}}
\def\herschel{{\em Herschel}}

\def\cm2{{cm$^{-2}$}}

\def\Hb{{H$\beta$}}
\def\Ha{{H$\alpha$}}
\def\oii{[\ion{O}{ii}]}
\def\oiii{[\ion{O}{iii}]}

\def\Hanii{{H$\alpha$}/[\ion{N}{ii}]}
\def\ciraca{\mbox{[3.6]--[4.5]}}
\def\ciracb{\mbox{[5.8]--[8.0]}}
\def\micron{$\mu$m}

\begin{document}

   \title{Multi-wavelength landscape of the young galaxy cluster RXJ1257.2+4738 at z=0.866}
   \subtitle{I. The infrared view\thanks{{\it Herschel} is an ESA space observatory with science instruments provided by European-led Principal Investigator consortia and with important participation from NASA.}}


   \author{I. Pintos-Castro
          \inst{1,2,3}
          \and
         M. S\'anchez-Portal \inst{4,5}
          \and
         J. Cepa \inst{2,1}
          \and
         J. S. Santos \inst{4}
          \and
         B. Altieri \inst{4}
          \and
         R. P\'erez Mart\'inez \inst{6,5}
          \and
         E. J. Alfaro \inst{7}
          \and
         \'A. Bongiovanni \inst{1,2}
          \and
         D. Coia \inst{4}
          \and
         L. Conversi \inst{4}
          \and
         H. Dom\'inguez-S\'anchez \inst{1,2}
          \and
         A. Ederoclite \inst{8} 
          \and
         J.I. Gonz\'alez-Serrano \inst{9}
          \and
         L. Metcalfe \inst{4}
          \and
         I. Oteo \inst{1,2}
          \and
         A.M. P\'erez Garc\'ia \inst{1,2}
          \and
         J. Polednikova \inst{1,2}
          \and
         T. D. Rawle \inst{4}
          \and
         I. Valtchanov \inst{4}
          }

   \institute{Instituto de Astrof\'isica de Canarias, La Laguna, Tenerife, Spain
              \\
              \email{ipc@iac.es}
         \and
             Departamento de Astrof\'isica, Facultad de F\'isica, Universidad de La Laguna, La Laguna, Tenerife, Spain
         \and
             Centro de Astrobiolog\'ia, INTA-CSIC, Villanueva de la Ca\~nada, Madrid, Spain 
         \and
         	Herschel Science Centre, ESAC/ESA, Villanueva de la Ca\~nada, Madrid, Spain
	     \and
	        ISDEFE, Madrid, Spain.
	     \and
	        XMM/Newton Science Operations Centre, ESAC/ESA, Villanueva de la Ca\~nada, Madrid, Spain
         \and
            Instituto de Astrof\'isica de Andaluc\'ia, CSIC, Granada, Spain
	     \and
	        Centro de Estudios de F\'isica del Cosmos de Arag\'on, Teruel, Spain
         \and
            Instituto de F\'isica de Cantabria (CSIC-Univ. de Cantabria), Santander, Spain
         }

   \date{Received month day, year; accepted month day, year}

  \abstract
   {Many studies have shown how galaxy properties (e.g. colours, morphology, star-forming activity, active galactic nuclei  population) change not only with redshift, but also with local galaxy density, revealing the important effect of the stellar/halo mass and the environment in the evolution of galaxies. A detailed analysis of the star formation activity in a representative sample of clusters will help us to understand the physical processes that cause the observed changes.}
   {We performed a thorough analysis of the star formation activity in the young massive galaxy cluster RXJ1257+4738 at z=0.866, with emphasis on the relationship between the local environment of the cluster galaxies and their star formation activity. We present an optical and infrared (IR) study that benefited from the large amount of data available for this cluster, including new OSIRIS/GTC and \herschel\ imaging observations.}
   {Using a multi-wavelength catalogue from the optical to the near-infrared , we measured photometric redshifts through a $\chi^2$ spectral energy distribution fitting procedure. We implemented a reliable and  carefully chosen cluster membership selection criterion including Monte Carlo simulations and derived a sample of 292 reliable cluster member galaxies for which we measured  the following properties: optical colours, stellar masses, ages, ultraviolet luminosities and local densities. Using the MIPS 24\,$\mu$m and \herschel\ data, we measured total IR luminosities and star formation rates (SFR).}
   {Of the sample of 292 cluster galaxies, 38 show far-infrared (FIR) emission with an SFR between 0.5 and 45\,M$_\odot$yr$^{-1}$. The spatial distribution of the FIR emitters within the cluster density map and the filament-like overdensities observed suggest that RXJ1257 is not virialised, but is in the process of assembly. The average star formation as a function of the cluster environment parametrised by the local density of galaxies does not show any clear trend. However, the fraction of star-forming (SF) galaxies unveils that the cluster intermediate-density regions is preferred for the SF activity to enhance, since we observe a significant increase of the FIR-emitter fraction in this environment. Focusing on the optically red SF galaxies, we cannot support the interpretation of this population as dusty red galaxies, since we do not observe an appreciable difference in their extinction compared with the quiescent or the blue populations.}
   {}

               
    \authorrunning{I. Pintos-Castro et al.}          
	\titlerunning{The IR view of RXJ1257.2+4738}
	\maketitle

	\section{Introduction}\label{sec:intro}
        Evidence has accumulated for decades, proving that the environment plays an important role in the history of galaxy evolution. The range of densities found in the surroundings of galaxy clusters make them ideal laboratories for studying the effects of local environment on their galaxies. In nearby clusters, \citet{Dressler1980} found that red early-type galaxies dominate in high-density regions, while blue late-type galaxies tend to live in low-density environments. At intermediate redshifts, between z=0.2 and z=0.5, the fraction of blue star-forming (SF) galaxies in clusters increases \citep[the Butcher-Oemler effect,][]{Butcher1984}, maintaining the fraction of ellipticals, but decreasing the fraction of S0 \citep{Dressler1997}. Many studies provided additional evidence of galaxy evolution: in distant clusters, infrared (IR) surveys have shown the increase of dusty star formation \citep{Coia2005,Geach2006,Saintonge2008}, and X-ray observations have found a growing population of active galactic nuclei \citep[AGN][]{Martini2002}. Focusing in a single epoch, changes in galaxy properties with clustercentric distance have been identified, such as gradients in colour \citep{Pimbblet2001} and spectral features \citep{Balogh1999}.
        
	   The morphology segregation can be understood as a consequence of a morphological transformation of spiral galaxies into S0, because they fall into high-density regions. Nevertheless, it is still unknown whether the quenching of the star formation activity is due to the advanced evolution in overdense regions, or to a direct physical effect on the star formation capability of galaxies in dense environments \citep{Popesso2007}. Possible mechanisms proposed to modify the star formation include \citep{Treu2003,Boselli2006} \textit{(i)} interactions of a galaxy with the gaseous component of the cluster (ram-pressure stripping, thermal evaporation of the interstellar medium (ISM), turbulent and viscous stripping, pressure-triggered star formation); \textit{(ii)} interactions of cluster galaxies with the cluster gravitational potential (tidal compression and tidal truncation); and \textit{(iii)} interactions between galaxies (mergers and harassment).
	   
	   To disclose the nature of the processes that produce the galaxy transformation in clusters we need measurements of star formation activity at different cosmic times across large radial distances and covering a wide range of cluster masses. Measuring the \Ha\ and \oii\ features in emission line galaxies (ELGs) is a method often used to trace star formation rates (SFRs) in clusters of galaxies \citep{Kodama2004,Finn2005,Koyama2010,Hayashi2010}, usually using narrow-band filters. However, these surveys are sensitive to metallicity and suffer from dust extinction. Observations with mid-infrared (MIR) and far-infrared (FIR) instruments have enabled the study of the dust-obscured star formation in many intermediate- and high-redshift clusters \citep{Koyama2008,Saintonge2008,Tran2010}, but these studies are hampered by the lack of spectroscopic confirmation of the emission galaxies.
	   
	  To overcome these difficulties, we are performing a complete study of the galaxy properties of the cluster RXJ1257.2+4738 (hereafter referred to as RXJ1257) at z=0.866 as part of the GaLaxy Cluster Evolution Survey (GLACE) \citet[][S\'anchez-Portal et al. 2013 in prep.]{SanchezPortal2011}. This project aims to study the evolution of the emission-line galaxies in clusters, using tunable filters (TF), in an initial sample of six clusters at redshifts 0.4, 0.63, and 0.86. The GLACE observations include broad-band imaging in the Sloan filters as well as narrow-band imaging performed with the TF of the OSIRIS\footnote{Optical System for Imaging and low Resolution Integrated Spectroscopy}\citep[][]{Cepa2003} instrument at the 10.4\,m GTC telescope at La Palma, mapping the \oii, \oiii\ and \Hb\ lines. TF imaging will allow us to determine accurate cluster membership, reaching a SFR down to $\sim$\,2\,M$_{\odot}$yr$^{-1}$ and to distinguish AGNs (Pintos-Castro et al., in preparation).  The galaxy cluster RXJ1257 was included in the GLACE sample because at the cluster redshift the lines  \oii, \oiii\ and \Hb\ fall within atmospheric windows relatively free of sky lines, and furthermore, this cluster has a wealth of ancillary data.
	  	  
	  The RXJ1257 cluster was first detected in X-rays in the ROSAT data archive, and the i$'$-band and Ks-band follow-up observations found a concentration of red galaxies that confirmed the existence of a cluster. \cite{Ulmer2009} reported the discovery of the spectroscopically confirmed cluster at z=0.866, on the basis of an extensive study including \textit{Chandra}, \textit{XMM-Newton}, \spitzer, Gemini, Subaru and ARC observations. They detected the diffuse emission of the cluster both in \textit{Chandra} and \textit{XMM-Newton} observations, and modelled that emission with an ellipse centered at RA(J2000)\,=\,12h\,57m\,12.2s Dec(J2000)\,=\,+47$^\circ$\,38$'$\,06.5$''$. The cluster X-ray luminosity is about 2\,$\times$\,10\,$^{44}$\,erg\,s$^{-1}$ and its temperature is about 3.6\,keV.  \cite{Ulmer2009} inferred an X-ray mass of (1\,-\,5)\,$\times$\,10$^{14}$\,M$_{\odot}$ and, based on a velocity dispersion of 600\,km/s, they obtained an independent mass estimate of 6.1\,$\times$\,10$^{14}$\,M$_{\odot}$. These mass values are consistent with RXJ1257 being a massive cluster. Based on the temperature of the intracluster medium (ICM), these authors derived the virial radius of the cluster as R$_V$\,=\,1.05\,Mpc.
	  
     The structure of this paper is as follows: in Section 2 we present the data sample, including the multi-wavelength catalogue generation. The selection of cluster candidates is explained in Section 3. The optical and infrared properties of the selected cluster members are shown in Section 4. In Section 5 we analyze the environmental dependence of the FIR cluster members. In Section 6 we discuss the dependence of the properties of the star-forming cluster galaxies with the environment, and in Section 7 we list the main conclusions of this work.

     Throughout this paper we assume a ﬂat Universe with H$_0$\,=\,70\,km\,s$^{-1}$\,Mpc\,$^{-1}$, $\Omega_{\Lambda,0}$\,=\,0.7 and $\Omega_{m,0}$\,=\,0.3. Magnitudes are listed in the AB system \citep{Oke1983}, unless indicated otherwise.

	\section{Observations, data reduction, and catalogue}
\begin{figure}
\centering
\includegraphics[width=\hsize]{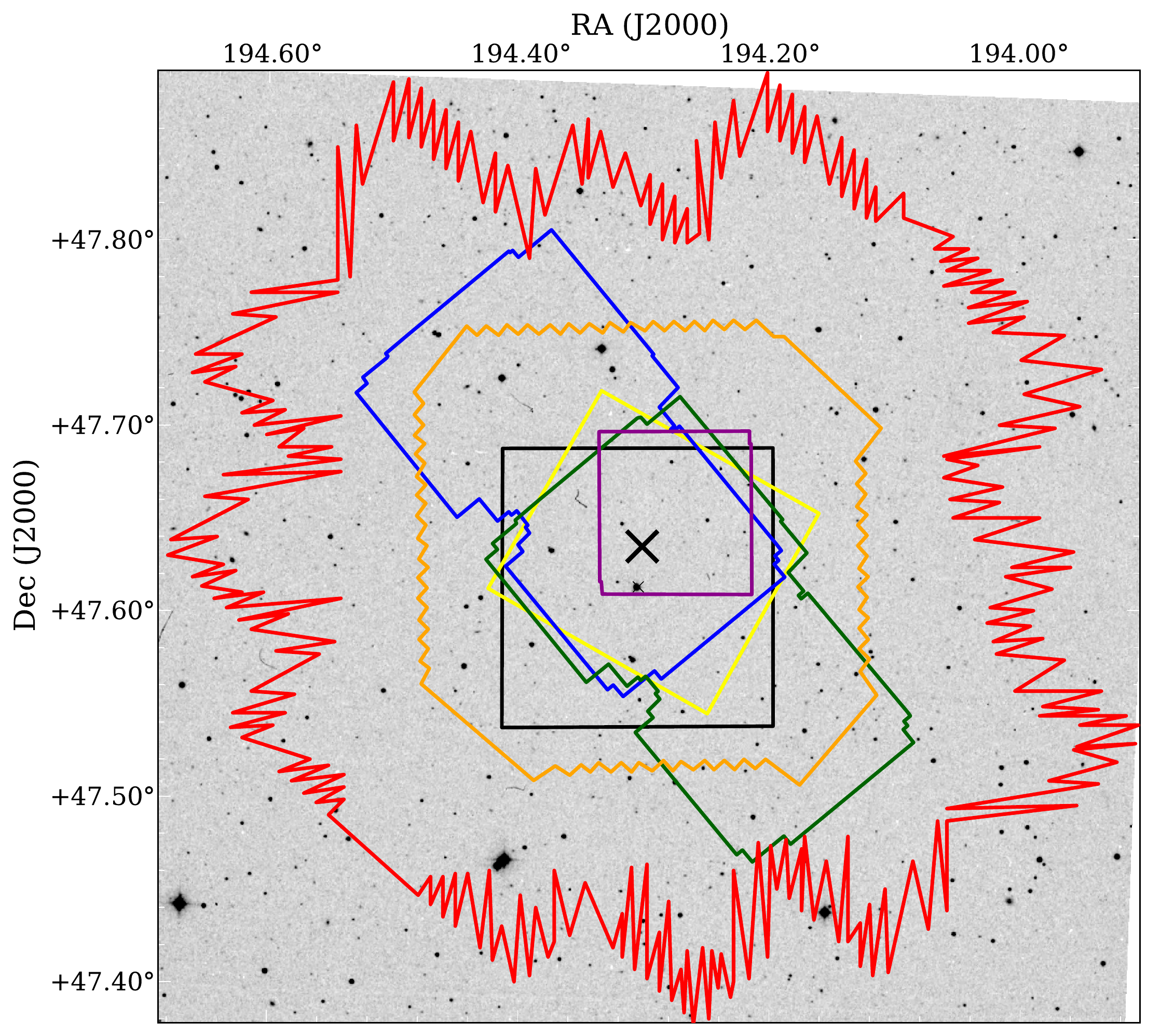}
\caption{Field of view of our observations superimposed on an $r'$-band Palomar Schmidt image. The black contour indicates the FoV of OSIRIS$/$GTC (the region analysed in this work covering 7.6$'$\,$\times$\,7.3$'$); purple: LIRIS$/$WHT (4.27$'$\,$\times$\,4.27$'$); blue: IRAC(3.6,5.8)$/$\spitzer\ (13.7$'$\,$\times$\,7$'$); green: IRAC(4.5,8)$/$\spitzer\ (13.7$'$\,$\times$\,7$'$); yellow: MIPS$/$\spitzer\ (7.4$'$\,$\times$\,8$'$); orange: PACS$/$Herschel (14.9$'$\,$\times$\,14.2$'$); red: SPIRE$/$\herschel\ (23.5$'$\,$\times$\,21.5$'$). The cross symbol ($\times$) marks the centroid of the X-ray emission. This cluster shows no clear dominant galaxy.}\label{fig:fov}
\end{figure}

We used optical and near-infrared (NIR) data of the GLACE project, FIR data from the  \herschel\ Space Observatory Guaranteed Time (GT) programme ``The star formation history of galaxy clusters'' (P.I. B. Altieri), and archival data from the Gemini Observatory (GMOS spectroscopy) and the \spitzer\ Space Telescope (IRAC and MIPS imaging). As can be seen in Fig.\,\ref{fig:fov}, the footprints of the observations are very heterogeneous, with the sky area covered by the OSIRIS/GTC images defining the region analysed in this study. This is the area where we estimate to find reliable cluster candidates, including most of the MIR/FIR emitters.

		\subsection{Optical observations}
			\subsubsection{Photometry}\label{subsec:optphotometry}
			Broad-band imaging with SDSS $g'r'i'z'$ filters was carried out using OSIRIS \citep{Cepa2003}, an imager and spectrograph for the optical wavelength range located at the Nasmyth-B focus of the 10.4\,m Gran Telescopio de Canarias (GTC). The detector consists of a mosaic of two Marconi CCD42-82 of 2048\,$\times$\,4096\,pixels with a 72-pixel gap between them, covering a total field of view (FoV) of 8$'$\,$\times$\,8$'$ (7.8$'$\,$\times$\,7.8$'$ unvignetted field). The pixel scale is 0.25$''$ in the standard mode of 2\,$\times$\,2 binning .
			
			The observations were obtained in service mode during 22 January and 28 March 2012 as part of a large ESO/GTC programme (ID. 186.A-2012, P.I. M. S\'anchez-Portal). The data were gathered in two dark nights under photometric conditions, with a seeing below 1.1$''$. Total exposure times were 3\,$\times$\,60\,s for $g'$, 3\,$\times$\,150\,s for $r'$ and $i'$, and 25\,$\times$\,80\,s for $z'$. Although we have designed the observations with a dithering pattern, we have not covered the gap area to avoid a bright star near the cluster center.
			
			The data were reduced using IRAF\footnote{IRAF is distributed by the National Optical Astronomy Observatory, which is operated by the Association of Universities for the Research in Astronomy, Inc., under cooperative agreement with the National Science Foundation (http://iraf.noao.edu/)} procedures. Individual frames were bias subtracted, flat-fielded, aligned, and coadded to produce deep images in each band. For the $z'$ images, additional fringe patterns were removed by subtracting a master fringe frame obtained by combining 3\,$\times$\,3 dithered science frames. Astrometric corrections were performed using the Sloan Digital Sky Survey Release 6 catalogue (SDSS-DR6), with \texttt{ccxymatch} and \texttt{ccmap} tasks. The astrometry accuracy is better than 0.17$''$. We transformed instrumental magnitudes into observed magnitudes with observations of photometric standard stars, obtained on 23 January 2012, by comparing them with SDSS-DR6. Calculated zero points are 28.31\,$\pm$\,0.02, 29.05\,$\pm$\,0.02, 28.25\,$\pm$\,0.02, and 27.72\,$\pm$\,0.03 for $g'$,$r'$,$i'$, and $z'$, respectively. We corrected for atmospheric and galactic \citep{Calzetti2000} extinctions.  
			
			Both source detection and photometry were performed with SExtractor \citep{Bertin1996}. The total flux of each source was measured using the SExtractor FLUX\_AUTO parameter. The $g'r'i'z'$ catalogues included objects with a detection threshold $>$\,2$\sigma$ in all bands, with a completeness level of 25.0, 24.5, 24.2, and 23.5\,mag for $g'$, $r'$, $i'$, and $z'$, respectively. The matched catalogue included 1985 sources (1962 $g'$-detected, 1979 $r'$-detected, 1963 $i'$-detected, and 1859 $z'$-detected).
			
			\subsubsection{Spectroscopy}\label{subsec:spectra}
			We reduced and analysed the available raw Gemini north GMOS \citep{Hook2004} observations to spectroscopically confirm 21 cluster members. The data were taken as part of three Gemini programmes, targeting 72 galaxies in total, each with a different mask and instrumental setup. In the GN-2005A-Q-9 programme 12 spectra were taken using multi-slit nod-and-shuffle (N\&S) mode, with 2$''$-wide slits. The mask was exposed 3\,$\times$\,4560\,s using the R400\_G5305 grating blazed at 7610\,\AA, 7640\,\AA\ and 7670\,\AA, with the detector binned 4\,$\times$\,4, providing a spectral resolution of 16\,\AA. The  MOS mask for the GN-2006A-Q-4 programme consisted of 35 slits of width 1.5$''$. This mask was exposed using the R400\_G5305 grating for 3\,$\times$\,2500\,s at a central wavelength of 7500\,\AA, with the detector binned 2\,$\times$\,2, producing a spectral resolution of 8\,\AA. For the GN-2006B-Q-38 programme the multi-slit N\&S mode was used to take 31 spectra with 1.5$''$-wide slits. The R400\_G5305 grating was used blazed at 7610\,\AA\ to expose the mask 5\,$\times$\,2560\,s with the detector binned 2\,$\times$\,2, resulting in a spectral resolution of 8\,\AA.
			
			The data were reduced using the Gemini/GMOS IRAF package. The standard routine was used to carry out bias subtraction, flat-fielding, cosmic ray removal, wavelength calibration, sky subtraction, and extraction into 1D single spectra. All spectra were examined interactively and compared with a variety of spectral line lists. Redshifts were derived by identifying a set of lines at a common redshift, fitting a Gaussian profile to calculate the central wavelength, and taking the average redshift from all detected lines. The majority of redshifts were determined by the \oii\ 3727\,\AA\ emission line and the Calcium II H+K absorption lines. In addition, several Balmer lines are visible in many spectra. From the total of 72 spectra, 31 were discarded because of  the poor quality of the data. Hence, for 41 spectra we measured redshifts between $\sim$\,0.19 and $\sim$0.89, of these 21 were selected as cluster members with redshift between 0.85 and 0.88, corresponding to a velocity dispersion $\sim$\,$\pm$\,2500\,km/s relative to the cluster redshift.
			
		\subsection{Near-infrared observations}
			J-band imaging was carried out on 12 and 13 May 2011 with the NIR imager/spectrograph LIRIS \citep[Long-Slit Intermediate Resolution Infrared Sectrograph,][]{Manchado2004,Acosta2003} on the William Herschel Telescope (WHT) as part of the Spanish CAT programme 197-WHT59/11A (P.I. M. S\'anchez-Portal). This instrument uses a Rockwell Hawaii 1024x1024 HgCdTe array detector, with a pixel scale of 0.25$''$, yielding a FoV of 4.27$'$\,$\times$\,4.27$'$. The observations were performed by adopting a dithering pattern that consisted of taking exposures in five different positions across the detector, starting at the centre, moving to the lower left quadrant and continuing with the remaining quadrants in a clockwise direction. We took three exposures of 60\,s in each of these five positions, and repeated the pattern four times each night giving a total exposure of two hours.
			
			The data were processed using the LIRISDR dedicated software\footnote{developed by Jos\'e Acosta Pulido, see \url{http://www.iac.es/galeria/jap/lirisdr/LIRIS_DATA_REDUCTION.html}} within the IRAF environment. Before any reduction procedure, we corrected the pixel-mapping anomaly of LIRIS with the \texttt{lcpixmap} task. The first step was flat-field correction of individual frames with \texttt{lmkflat} and \texttt{liflatcor}. Then, the complete task \texttt{ldedither} was executed. This task computes a sky subtraction, a vertical gradient, and a geometrical distorsion correction to finally shift and combine all frames into a single final image.
			
			Photometric calibration was made by estimating zero points by comparing the fluxes of 63 bright sources (selected from five different fields observed during the two nights) with their magnitudes in the 2MASS All-Sky Catalogue. We did not correct for atmospheric extinction because it is negligible at these wavelengths. Astrometry was corrected as in section \ref{subsec:optphotometry}, and we achieved an accuracy below 0.19$''$, which is lower than the pixel scale. 
			
			We constructed a catalogue in the J-band with SExtractor, using FLUX\_AUTO for total fluxes. To avoid spurious detections we set a detection threshold of 3$\sigma$, obtaining a sample of 947 sources whose limiting magnitude is around 23.5. For many galaxies J photometry was not available, due to the small area covered by these observations, but this data point is not critical for the determination of the main parameters  (i.e. redshifts, stellar masses, IR luminosities).
			
		\subsection{Mid-infrared observations}
		The \spitzer\ obsevations of RXJ1257 were obtained with the Infrared Array Camera \citep[IRAC,][]{Fazio2004} on 29 December 2006 and with the Multiband Imaging Photometer for Spitzer \citep[MIPS,][]{Rieke2004} on 2 June 2007. These IRAC and MIPS archival images were obtained from the Spitzer public archive. The IRAC observations consisted of 20 frames with an integration time of 100\,s in each of the four channels centered at 3.6, 4.5, 5.8, and 8\,$\mu$m. The FoV of the different IRAC channels did not fully overlap: there are two adjacent fields in the focal plane viewed by the channels in pairs (3.6 and 5.8\,$\mu$m; 4.5 and 8\,$\mu$m). The MIPS observations were taken in the photometry mode in the 24\,$\mu$m channel with 30\,s integration time per frame, resulting in a total exposure time of 9050\,s.
		
		The IRAC corrected basic calibrated data (CBCD) include corrections for column pulldown and banding and remained sufficiently clean for the analysis. We processed these CBCD images to compute overlap correction and to produce a photometrically corrected science mosaic using MOPEX \citep{Makovoz2005}. The MIPS 24\,$\mu$m mosaic exhibited artefacts, particularly jailbars, and gradients. To remove the artefacts we used a self-calibration script which performs background subtraction, source detection, and coaddition of background-subtracted images to create a flat and flat-fielding of the basic calibrated data (BCD) images. Finally, we created the mosaic using our self-calibrated BCD.
		
		To extract IRAC fluxes we used APEX (the point-source extraction package of MOPEX), performing detection on co-added data, and point-source fitting photometry on the individual frames. We ran APEX with the default parameters and applied a small flux correction to account for differences in normalisation between the IRAC data and APEX. The extracted catalogues included 2170, 2191, 2057, and 1085 objects, reaching completeness levels of 22.5, 22.0, 20.5, and 20.5\,mag for the 3.6, 4.5, 5.8, and 8\,$\mu$m bands, respectively. MIPS 24\,$\mu$m fluxes were measured by combining SExtractor and APEX utilities. We detected and extracted sources running SExtractor with a detection limit of 3$\sigma$, finding that some clear sources were lost due to inaccurate deblending. Hence, we made an auxiliary list of sources and extracted them using the ``APEX User List" module. We obtained a catalogue of 297 sources complete up to 19.2\,mag for the  24\,$\mu$m band.
		
		\subsection{Far-infrared observations}\label{subsec:firobs}
		
		The \herschel\ maps using the PACS photometer at 100 and 160\,$\mu$m and the SPIRE photometer at 250, 350 and 500\,$\mu$m were observed on  28 May 2011 and 22 November 2010, respectively. The PACS observations were acquired in prime mode and consist of two crossed-square maps of 9075\,s each at the nominal scan speed of 20\,arcsec/s. The scan maps had 30 legs of 10\,arcmin length, separated by 20\,arcsec, and the pattern was repeated six times. The SPIRE bands, which reach confusion relatively fast, have 2047\,s total integration time per channel. PACS data were processed in the HIPE environment \citep{Ott2006}. The data cubes were processed with a standard pipeline, including a sliding high-pass filtering on the detector time lines to remove detector drifts and 1/f noise with an iterative masking of the sources.  We extracted flux densities via standard aperture photometry techniques, and the PACS uncertainties include contributions from instrumental noise and absolute flux calibration. The projected maps have a pixel size of 2$''$ at 100\,$\mu$m and 3$''$ at 160\,$\mu$m.

    For the PACS 100 and 160\,$\mu$m data we produced blind-detection catalogues using SExtractor. We performed aperture photometry considering a radius of 6$''$ in the 100\,$\mu$m channel and 9$''$ in the 160\,$\mu$m band, which were then corrected to compensate for flux loss outside that area using the tabulated values in \citet{Mueller2011}. The 3$\sigma$ sensitivity limits are 4.2\,mJy at 100\,$\mu$m and 9.0\,mJy at 160\,$\mu$m. 

     SPIRE fluxes were estimated using SUSSEXtractor \citep{Smith2012}. For SPIRE 250\,$\mu$m we compiled a blind catalogue with a detection threshold of 3$\sigma$. For SPIRE 350 and 500\,$\mu$m we used the 250\,$\mu$m prior positions. For FIR undetected sources we considered  the following upper limits: 3.0, 6.0, 13.4, 15.7, and 18.8\,mJy for 100, 160, 250, 350, and 500\,$\mu$m, respectively. The catalogues included 212, 160, 614, 543, and 443 sources in each band. The total FIR luminosity at 3$\sigma$ is estimated to be 2.4\,$\times$\,10$^{11}$\,L$_\odot$.
				
		\begin{table}[]
	    \caption{Summary of datasets.}\label{tab:data}
		\begin{center}
		\begin{tabular}{l c r@{  /  }l c}
		\hline\hline
		\multirow{2}{*}{\textbf{Instrument}} & \multirow{2}{*}{\textbf{Band}} & \multicolumn{2}{c}{\textbf{N$^{\circ}$ of sources}} & \multirow{2}{*}{\textbf{Completeness}} \\
		&  & \textbf{Total} & \textbf{Region}\tablefootmark{a} & \\
		\hline
		OSIRIS & $g'$ & 1962 & 1962 & 25.0\,mag \\
		& $r'$ & 1979 & 1979 & 24.5\,mag \\
		& $i'$ & 1963 & 1963 & 24.2\,mag \\
		& $z'$ & 1859 & 1859 & 23.5\,mag \\
		\hline
		LIRIS & J & 947 & 947 & 23.5\,mag \\
		\hline
		IRAC & 3.6\,$\mu$m & 2170 & 1019 & 22.5\,mag \\
		& 4.5\,$\mu$m & 2191 &  1128 & 22.0\,mag \\
	    & 5.8\,$\mu$m & 2057 &  848 & 20.5\,mag \\
		& 8\,$\mu$m & 1085 &  550 & 20.5\,mag \\
		\hline
		MIPS & 24\,$\mu$m & 297 &  250 & 19.2\,mag \\
		\hline
		PACS & 100\,$\mu$m & 212 &  81& 4.2\,mJy \\
		& 160\,$\mu$m & 160 &  62 & 9.0\,mJy  \\
		\hline
		SPIRE & 250\,$\mu$m & 614 &  74 & 17.4\,mJy \\
		& 350\,$\mu$m & 543 &  69 & 18.9\,mJy \\
		& 500\,$\mu$m & 443 &  56 & 20.4\,mJy \\
		\hline
		\end{tabular}
		\tablefoot{\tablefoottext{a}{Number of sources in the region that correspond with the FoV of OSIRIS/GTC images}}	
		\end{center}
		\end{table}

		\subsection{Optical/near-infrared catalogue}
		With the aim of building a multi-wavelength catalogue that includes all detected sources, we cross-matched all individual catalogues, from optical to NIR, using the nearest-neighbour technique. With the optical catalogue as reference, we carried out consecutive matches towards longer-wavelength catalogues, associating the position of the shortest wavelength (typically the g$'$-band) for each source. The catalogue includes 1979 sources detected in any optical band. Following the methodology of \citet{Ruiter1977}, we estimated maximum-error radii of 2$''$ for optical and J-band catalogues and 3$''$ for the IRAC bands and made a quantitative analysis of our procedure, obtaining completeness and reliability values higher than 98\% and 82\%, respectively. This catalogue was used to determine the photometric redshifts.
		
		After defining the cluster-candidate sample using the photometric redshifts (see the next section), we proceeded to include the MIR/FIR data (see Sect.\,\ref{subsec:firsample}).

	\section{Determining the cluster members}
	\label{sec:clusterdet}
	In addition to the 21 spectroscopic cluster galaxies, we built a robust sample of photometrically selected cluster candidates following the selection steps outlined in the next subsections in the area enclosed by the OSIRIS field of view.
	\subsection{Spectral energy distribution fitting}\label{subsec:zphot}
\begin{figure}
\centering
\includegraphics[width=\hsize]{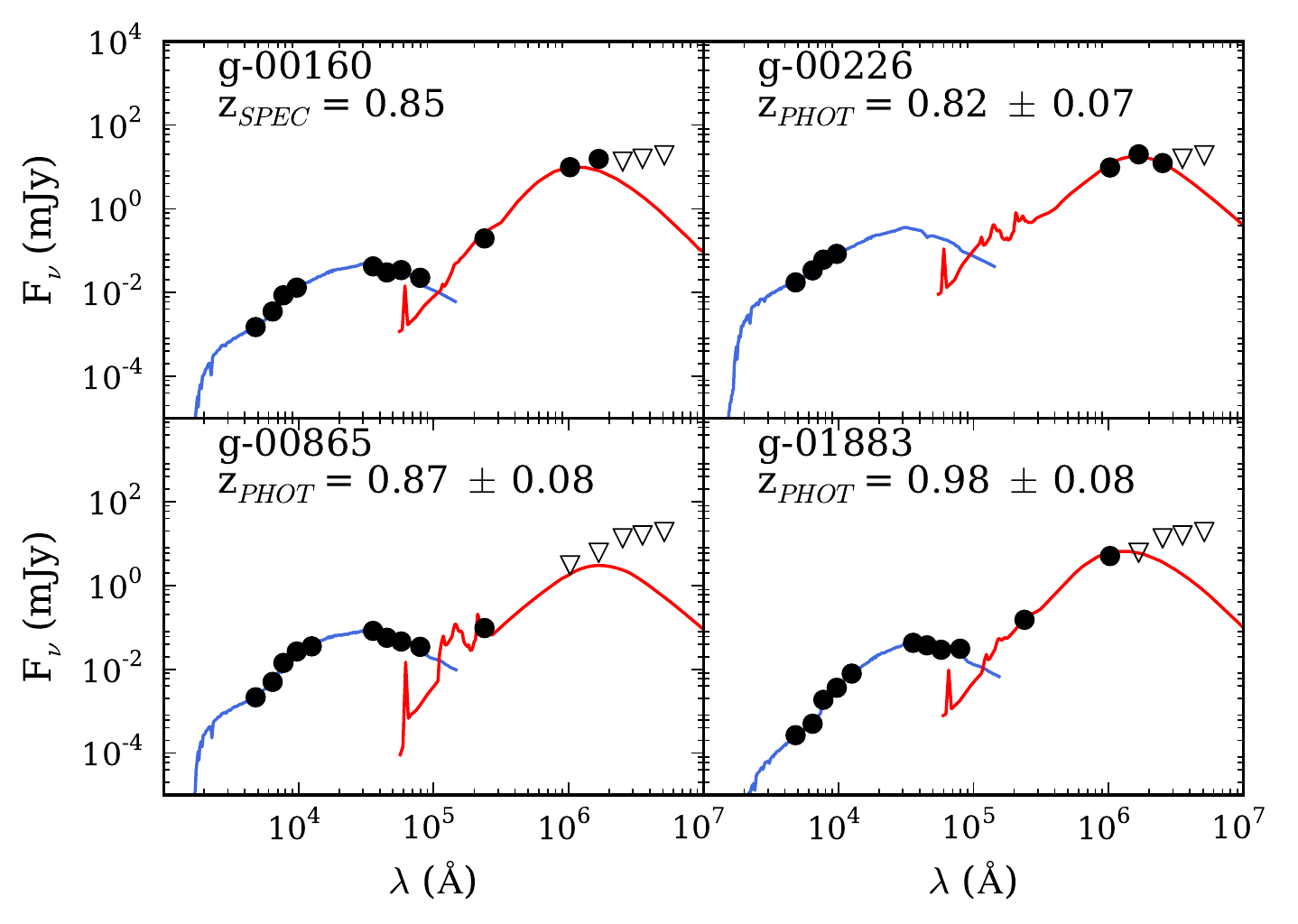}
\caption{Example of SED fitting for a representative sample of the FIR cluster galaxies: one spectroscopic cluster member, one Herschel-only detected, one MIPS-only detected and one detected in both MIPS and Herschel. Black filled circles show the observed photometry of each source, open triangles show upper limits. Blue curves represent the BC03 template that best fit the stellar part of the spectrum from the optical to IRAC 8\,$\mu$m. Red curves are the CE01 templates that best fit the fluxes from IRAC 8\,$\mu$m to SPIRE 500\,$\mu$m.}\label{fig:SEDsample}
\end{figure}

		Using our optical/NIR catalogue, we computed the photometric redshifts with the \textit{LePhare} code \citep{Ilbert2006}. \textit{LePhare} computes photometric redshifts based on a $\chi^2$ spectral energy distribution (SED) template-fitting procedure. To perform the SED-fitting we used the \citet[][hereafter BC03]{Bruzual2003} population synthesis models, with star formation histories exponentially declining with time as $e^{-t/\tau}$, to fit the stellar part of the spectrum. We included the bands between $g'$ and IRAC\,8\,$\mu$m. The complete template library was built considering three different metallicities (0.2\,Z$_{\odot}$,0.4\,Z$_{\odot}$, and Z$_{\odot}$) and nine different values of $\tau$ (0.1, 0.3, 1.0, 2.0, 3.0, 5.0, 10.0, 15.0, and 30.0 Gyr) with 57 steps in age, and applying the \cite{Calzetti2000} extinction law with ten different values of E(B-V) (0.0, 0.1, 0.2, 0.3, 0.4, 0.5, 0.6, 0.7, 0.8, and 0.9). We assumed the initial mass function (IMF) of \citet{Chabrier2003} and constrained the derived age of each galaxy to be younger than the age of the Universe at the galaxy redshift. In Fig. \ref{fig:SEDsample} some examples of the star-forming cluster member SED-fittings are shown.
		
	    \subsection{Cluster galaxies selection}\label{subsec:sim}
		To determine the most adequate photometric redshift range to provide a robust set of cluster member candidates we carried out Monte Carlo simulations. We created 600 mock catalogues by randomly varying the flux of each source in each band within a Gaussian distribution with a deviation proportional to the flux errors. Since the 1$\sigma$ flux errors obtained with SExtractor are unrealistically small, we assigned as error a percentage of the flux depending on the detection threshold of each source. Thus, for sources with a detection threshold up to 3$\sigma$ the deviation is 30\% of the flux, for sources with 4$\sigma$ it is 25\%, for sources with 5$\sigma$ it is 20\%, and for sources with a detection threshold down to 6$\sigma$ the deviation is 10\% of the flux. After running \textit{LePhare} for each mock catalogue, we obtained 600 photometric redshift distributions with a clear peak at the cluster redshift. We estimated the photometric redshift range of the cluster by fitting a Gaussian function. Each photometric redshift distribution was smoothed with a Gaussian kernel to fit a double Gaussian function to the probability density function. The second Gaussian function was necessary because of a second peak around a redshift of 2, which increases the right wing of the first Gaussian function shifts its centre. Considering the parameters of the first Gaussian function fitted, we defined the photometric cluster range to be 1$\sigma$ wide centred at the average central position, i.e. 0.886\,$\pm$\,0.185.

\begin{figure*}
\centering
\includegraphics[width=0.95\hsize]{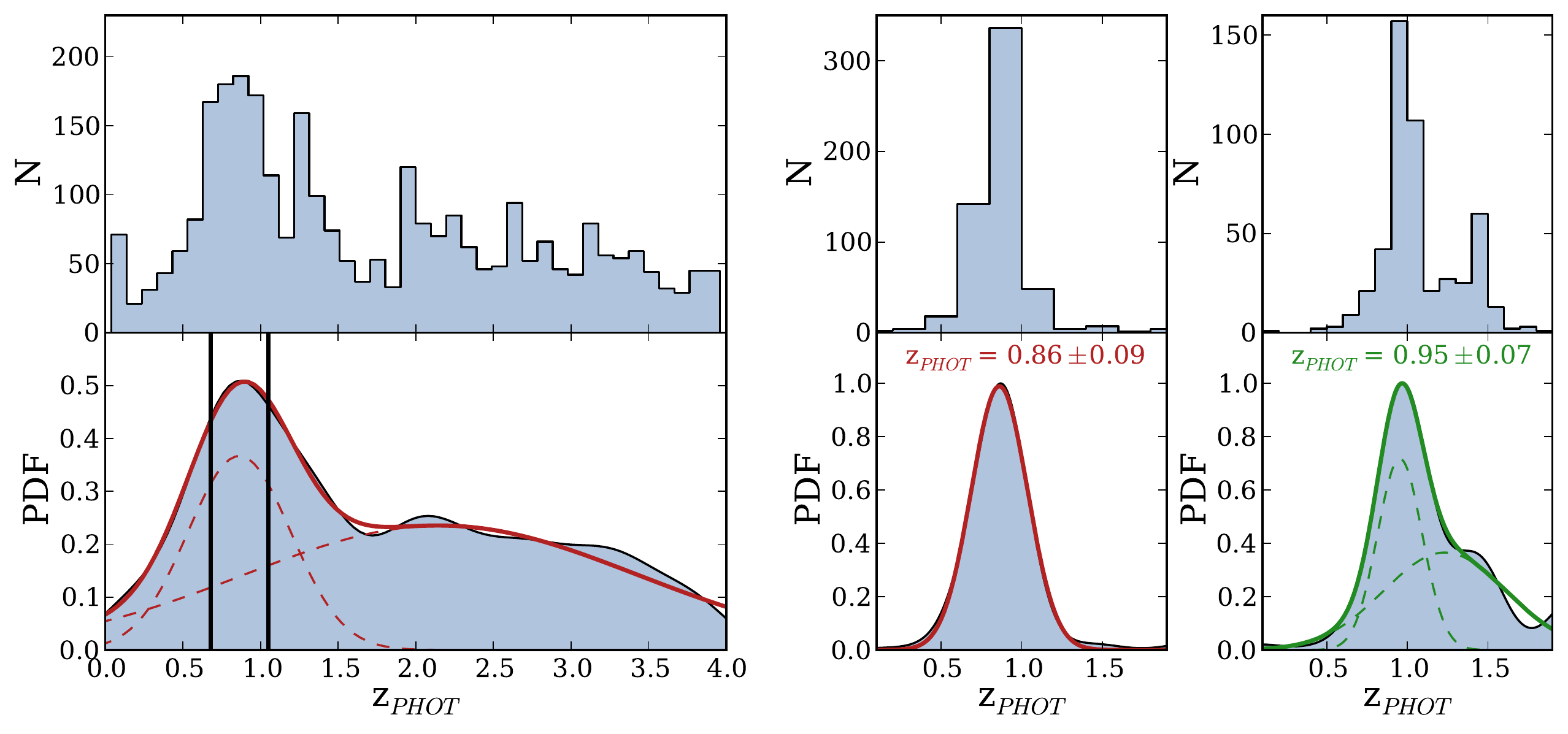}
\caption{Photometric redshift histograms (\textit{top}) and their probability density distributions (\textit{bottom}) for one Monte Carlo simulation (\textit{left}), one cluster galaxy fitted with a simple Gaussian (\textit{centre}), and an other cluster galaxy fitted with a double Gaussian (\textit{right}). Solid-coloured curves show the best fit obtained in each case, and dashed coloured curves show the individual Gaussian function for the double Gaussian fits. Black vertical lines indicate the locus of the photometric cluster range z$_{PHOT}$\,$=$\,[0.70, 1.07].}\label{fig:simulations}
\end{figure*}

\begin{figure}
\centering
\includegraphics[width=0.9\hsize]{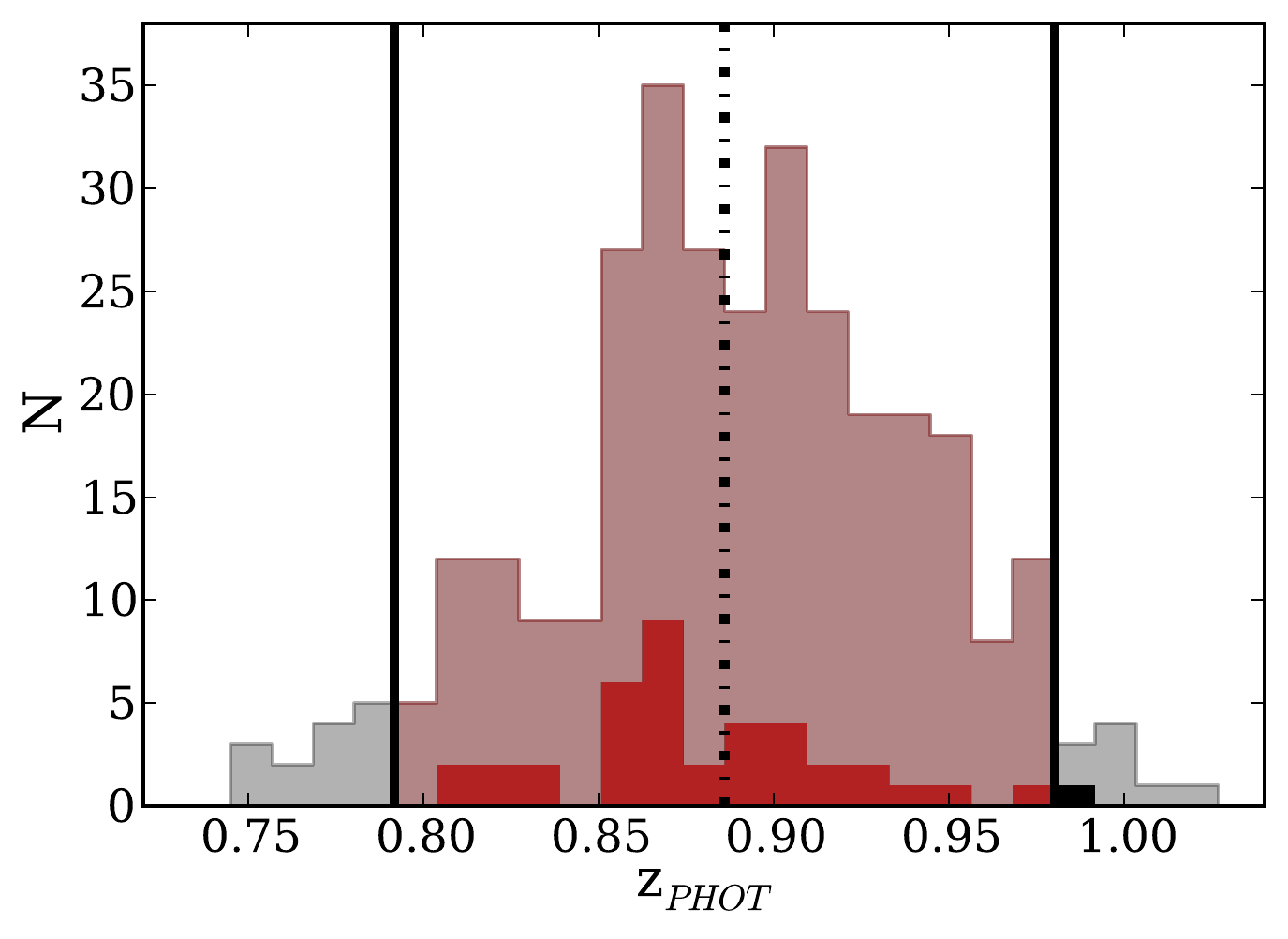}
\caption{Histogram of the redshifts obtained for the initial candidate sample (black) and for the final sample of 292 robust cluster candidates, 21 spectroscopic and 271 photometric (red). The filled histogram represents the FIR-emitter sample. The dot-dashed line at 0.886 indicates the locus of the photometric redshift of the cluster, and the solid lines shows the final photometric range we have considered.}\label{fig:zphothist}
\end{figure}

		Furthermore, to perform a refined selection of robust cluster member candidates we used the photometric redshift distribution for each object resulting from the set of 600 simulations. In a similar way to the procedure of determining the redshift range, we applied a Gaussian kernel to smooth the distribution and fitted a Gaussian function. When the probability density distribution of photometric redshifts had two peaks, a double Gaussian was fitted instead. The criterion we used to consider a source as a cluster candidate is that a 1$\sigma$ width centred at the central position of the best-fit Gaussian is included in the cluster redshift range. When a double Gaussian was needed we considered only the highest peak. Fig.\,\ref{fig:simulations} shows examples of the photometric redshift distributions obtained for one mock catalogue and two galaxies selected by the method as cluster members. With this criterion we extracted an initial candidate sample of 317 galaxies. As shown in Fig.\,\ref{fig:zphothist}, if we consider only the photometric redshift value of each source, the actual photometric redshift range is z$_{PHOT}$\,$=$\,[0.73, 1.02]. Additionally, we visually inspected the sample to compare the appearance of the photometric sample with the spectroscopically confirmed cluster members, which made us discard one foreground source because of its angular size.

		To test the quality of the photometric redshifts, we compared the derived photometric values with the sample of spectroscopic redshifts  available. From the 41 spectra obtained, see Sect.\,\ref{subsec:spectra}, only 35 have the optical photometry needed to derive reliable photometric redshifts (20 cluster members and 15 interloper galaxies). We estimated the redshift accuracy as the median of $\sigma_{\Delta z/(1+z_{SPEC})}$, where $\Delta z$\,=\,$z_{PHOT}$\,-\,$z_{SPEC}$, and defined the percentage of catastrophic errors as the sources with $|\Delta z|$\,/\,(1+$z_{SPEC}$)\,$>$\,0.15. The quantitative estimation of our method results in an accuracy of $\sigma_{\Delta z/(1+z_{SPEC})}$\,=\,0.05 and a percentage of catastrophic errors of 34\%. We emphasize that the selection method of the cluster members based on the photometric redshifts is optimised for the cluster redshift, therefore, we obtained an accuracy of $\sigma_{\Delta z/(1+z_{SPEC})}$\,=\,0.04 when we considered only the spectroscopic cluster sources  and none of the photometrically selected sources have $|\Delta z|$\,/\,(1+$z_{SPEC}$)\,$>$\,0.15.
		
		The mean error of the cluster galaxies $z_{PHOT}$, calculated with the $\sigma$ of the fitted Gaussian, is 0.08. This value corresponds to $\sigma_{\Delta z/(1+z_{SPEC})}$\,$\sim$\,0.045 at the cluster redshift $z$\,=\,0.866, which is consistent with the median of $\sigma_{\Delta z/(1+z_{SPEC})}$ derived from the comparison of the spectroscopic and photometric redshift cluster sample.
		
	   Finally, we defined a narrower photometric redshift range using the estimated accuracy at the cluster redshift. The value of $\sigma_{\Delta z/(1+z_{SPEC})}$ results in $\sim$\,0.094 at the photometric cluster central position,  which corresponds to a photometric redshift range of z$_{PHOT}$\,$=$\,[0.79, 0.98] (see vertical lines in Fig.\,\ref{fig:zphothist}). The final sample includes 292 robust cluster candidates, of which 21 are spectroscopic members.

       \subsection{Far-infrared cluster sample}\label{subsec:firsample}
        To build the FIR-emitter cluster sample, we cross-matched this full sample of 292 robust cluster candidates with the MIR/FIR catalogues using the nearest-neighbour technique. Following the methodology of \citet{Ruiter1977},  we considered a maximum-error radius of 4$''$ for the MIPS and PACS catalogues and 6$''$ for the SPIRE bands. The initial FIR-matched catalogue contains 67 sources. Due to the poor accuracy of the criterion which is based only on distance for the FIR counterparts, we made a visual inspection and analysed thumbnails of r$'$, IRAC 8\,$\mu$m, MIPS, PACS, and SPIRE bands, for each of these sources. From the initial sample of 67 FIR emitters, we discarded 15 sources because of confusion or blending problems, and 14 sources did not match  the longest wavelengths of \herschel. As a result, we obtained a reliable FIR cluster sample of 38 galaxies, 32 sources MIPS-only detected and one \herschel-only detected (this source is outside the MIPS FoV).


	
	\section{Optical and infrared galaxy properties}
    
	    \subsection{Optical colours of cluster galaxies}
	    \label{subsec:optcol}
\begin{figure}
\centering
\includegraphics[width=\hsize]{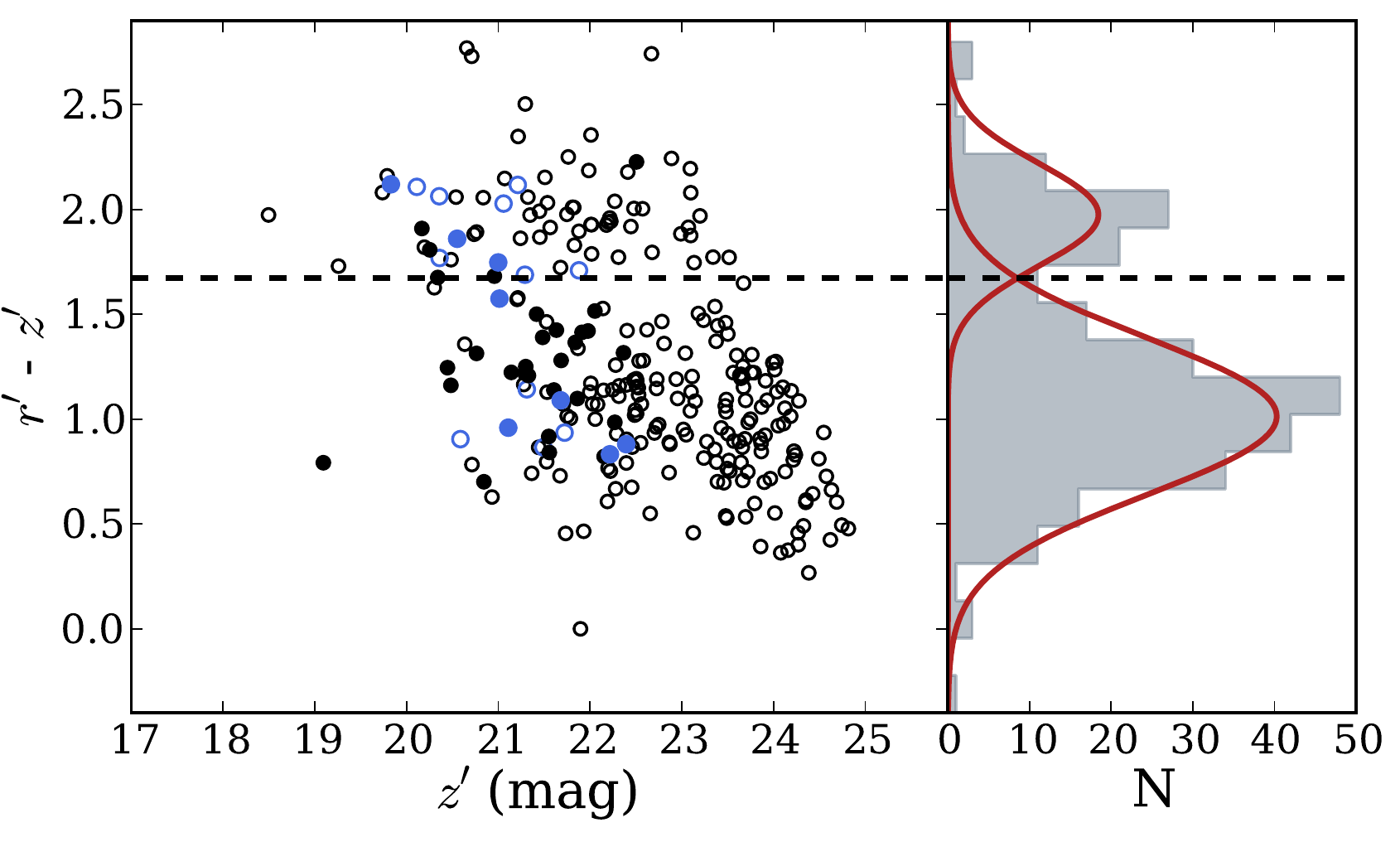}
\caption{Colour-magnitude diagram $r'$\,-\,$z'$ vs. $z'$ (\textit{right}) and the $r'$\,-\,$z'$ colour distribution (\textit{left}) of the cluster members. Open circles show the cluster members sample and filled circles are the FIR-emitter cluster sample. Blue circles are spectroscopic sources, and black circles are photometric cluster candidates. The red curves show the best-fit Gaussian functions and the dot-dashed black line indicates the boundary between optical red and blue galaxies.}\label{fig:redseq}
\end{figure}

	    Early-type cluster galaxies are known for the tight correlation between colour and absolute magnitude \citep{Visvanathan1977,Tully1982,Bower1992a}. \citet{Bell2004} also found that rest-frame colours for field galaxies are bimodal at all redshifts up to z\,$\sim$\,1. We used this colour bimodality to define a threshold between optically red and blue galaxies. Figure\,\ref{fig:redseq} shows the colour-magnitude diagram (CMD) of the cluster members (colours were calculated in the observed frame, i.e. no K-correction was applied), and their $r'$\,-\,$z'$ colour distribution. Although the CMD merely shows a tentative red sequence, the bimodality of the colour distribution is clearly seen. Therefore, fitting a Gaussian function to each colour peak, we defined the boundary between optical red and blue galaxies as the intersection point between the two curves. The resulting threshold is $r'$\,-\,$z'$\,=\,1.67.
	    
	    \subsection{Properties derived by optical SED-fitting}
	    \label{subsec:sedfitting}
	    As mentioned in Sect.\,\ref{subsec:zphot}, we used the \textit{LePhare} code to fit our multi-wavelength data with a set of optical/NIR SED templates from BC03. These models are generated with well-known physical properties, therefore the resulting best-fit SED contains all the information about the stellar population of that particular galaxy. \textit{LePhare} is designed to extract the physical information on individual galaxies and to provide several parameters as an output. We fixed the redshift to the spectroscopic value for the 21 spectroscopic cluster members and to the central position of the best-fit Gaussian for the 271 photometric cluster candidates. Fixing the redshift for each source, we executed \textit{LePhare} with the BC03 template library to obtain the best-fit stellar masses, ages, extinctions, and UV luminosities for the cluster sample.
	    
	    We can estimate the 1$\sigma$ uncertainties of these property as half the difference between the properties values for $\Delta\chi^{2}$\,=\,1 in their probability distribution. Using the \textit{LePhare} code, these values are labelled \_INF and \_SUP. The cluster galaxies cover the range of stellar masses M$_{\star}$\,=\,[\,6.7\,$\times$\,10$^{7}$, 8.8\,$\times$\,10$^{11}$\,]\,M$_{\odot}$, and the range of UV luminosities L$_{UV}$\,=\,[\,9.7\,$\times$\,10$^{6}$, 3.8\,$\times$\,10$^{11}$\,]\,L$_{\odot}$.
	
		\subsection{Infrared luminosities and star formation rates}
		\label{subsec:irlum}
		To fit the IR part of the SED we used the templates of \citet[][hereafter CE01]{Chary2001}. \textit{LePhare} allows us to fit the optical/NIR part of the spectra separately from the MIR/FIR part and provides the total infrared luminosity (L$_{IR}$), integrating the SED from 8 to 1000\,$\mu$m (rest frame). Of the 38 FIR emitters, 32 have been detected only in the MIPS 24\,$\mu$m images, 1 only in \herschel\ bands, and the remaining 5 sources have both MIPS and \herschel\ detections. For the 32 MIPS 24\,$\mu$m-only emitters, the total IR luminosity was  also computed using a single-band 24\,$\mu$m to L$_{IR}$ extrapolation as described in CE01, verifying that the two methods produce compatible results. Since there are only 5 FIR emitters with both MIPS 24\,$\mu$m and \herschel\  detections, a correction for the total IR luminosity  could not be estimated for the MIPS 24\,$\mu$m-only sources.
		
		The star formation rate is derived from the L$_{IR}$ according to the \cite{Kennicutt1998} relation modified for the Chabrier IMF \citep{Chabrier2003,Erb2006},
		\begin{equation}
        \rm SFR(M_{\odot}\,yr^{-1})\,=\,2.5\,\times\,10^{-44}L_{IR}\quad(erg\,s^{-1}).
		\end{equation}
		This simple conversion is consistent with the SFRs estimated by scaling the extinction-corrected \Ha\ flux, as shown by \citet{DominguezSanchez2012}. Thus, we calculated the specific star formation rate (sSFR\,=\,SFR\,/\,M$_{\star}$) in a coherent way.
		
		The 38 FIR cluster galaxies cover the range of IR luminosities L$_{IR}$\,=\,[5.3\,$\times$\,10$^{9}$, 4.7\,$\times$\,10$^{11}$]\,L$_{\odot}$, which corresponds to a range of star formation rates of SFR\,=\,[0.5, 45]\,M$_{\odot}$yr$^{-1}$. For the six FIR emitters detected in \herschel, the IR luminosities are higher than 1.7\,$\times$\,10$^{11}$\,L$_{\odot}$, and the corresponding SFRs exceed 16\,M$_{\odot}$yr$^{-1}$. These lower limits are compatible with the 3$\sigma$ L$_{IR}$ threshold estimated for the \herschel\ data in Sect.\,\ref{subsec:firobs}.
		
		The peak of the IR luminosity distribution is located at  $\sim$\,5.2\,$\times$\,10$^{10}$\,L$_{\odot}$ ($\sim$\,5.0\,M$_{\odot}$yr$^{-1}$), implying that below this IR luminosity the sample is probably incomplete. Of 38 FIR emitters, 19 are above this limit and 12 are in the luminous infrared galaxies range (LIRGs). Because of the reduced number of FIR-cluster candidates, we eventually decided to work with the full sample of 38 objects and kept in mind that the outcome may be affected by the lack of completeness.
				
		Using the UV luminosity obtained with the SED-fitting method and the \cite{Kennicutt1998} calibration, we estimated the UV contribution to the SFR.  With a mean contribution of 3\%,  we considered that the SFR$_{UV}$ is negligible and defined the total SFR as the SFR derived from the total IR luminosity. This consideration implies that for this study only galaxies with FIR emission are defined as star-forming galaxies. We realize that the cluster candidates not detected in the FIR could be forming stars with rates below our completeness limit of $\sim$\,5.0\,M$_{\odot}$yr$^{-1}$, but for simplicity, we consider the FIR-emitter sample as star-forming galaxies and the remaining cluster candidates as quiescent galaxies hereafter .


	Many studies have reported the correlation between the SFR and the stellar mass of star-forming galaxies over a wide range of redshifts \citep{Brinchmann2004,Noeske2007,Elbaz2007,Daddi2007}. \citet{Elbaz2011} have assumed a slope for this correlation of 1 at all redshifts to define a main-sequence mode of star formation with the median sSFR at a given redshift. In the right panels of Fig.\,\ref{fig:sfrssfrall} we show the SFR and sSFR of the FIR emitters as a function of the stellar mass. In the upper panel we include the SFR-M$_{\star}$ correlation fitted by Eq.(4) of \citet{Elbaz2007} using GOODS field data. In the lower panel we show the sSFR$_{MS}$ for RXJ1257 estimated with Eq.(13) of \citet{Elbaz2011}, considering that the age of the Universe at the cluster redshift is 6.32\,Gyr. The region above the dash-dotted line corresponds to the sSFR region of starburst, as defined by \citet{Elbaz2011}. Both panels show that almost all SF galaxies fall in or below the regime of main-sequence galaxies.

	  \subsection{AGN contamination}
			
		As supported by previous studies \citep[][and references therein]{Atlee2011}, it is expected that a significant fraction of the IR emission from galaxies within the cluster is produced by AGN activity instead of SF processes. The authors of the GLACE tunable filter \Hanii\ survey of the intermediate-redshift cluster Cl0024.0+1652 at z\,$=$\,0.395 (S\'anchez-Portal et al., 2013 in prep.; \citet{PerezMartinez2012}) derived that $\sim$\,20\% of emission-line galaxies (ELG) host an AGN by means of  a [NII]/H$\alpha$ vs. EW(H$\alpha$) diagnostic diagram \citep{CidFernandes2010}. At higher redshift, \citet{Lemaux2010} investigated the nature of the \oii\ emission in the RX J1821.6+6827 cluster at z\,$\sim$\,0.82  and in the  Cl1604 supercluster at z\,$\sim$\,0.9 (i.e. a redshift range comprising that of RXJ1257), finding that at least  $\sim$\,20\% of galaxies with M$_*$\,$>$\,10$^{10}$--10$^{10.5}$\,M$_{\odot}$ contain a LINER/Seyfert component that can be revealed with line ratios. In fact, within the framework of the on-going GLACE TF emission-line survey of RXJ1257 (Pintos-Castro et al., in prep., see also section \ref{sec:intro} above), we intend to use the diagnostic proposed by \citet{Rola1997} based on the equivalent widths of \oii\ and \Hb\ lines to properly address the AGN contents of the cluster.  
		
\begin{figure}
\centering
\includegraphics[width=\hsize]{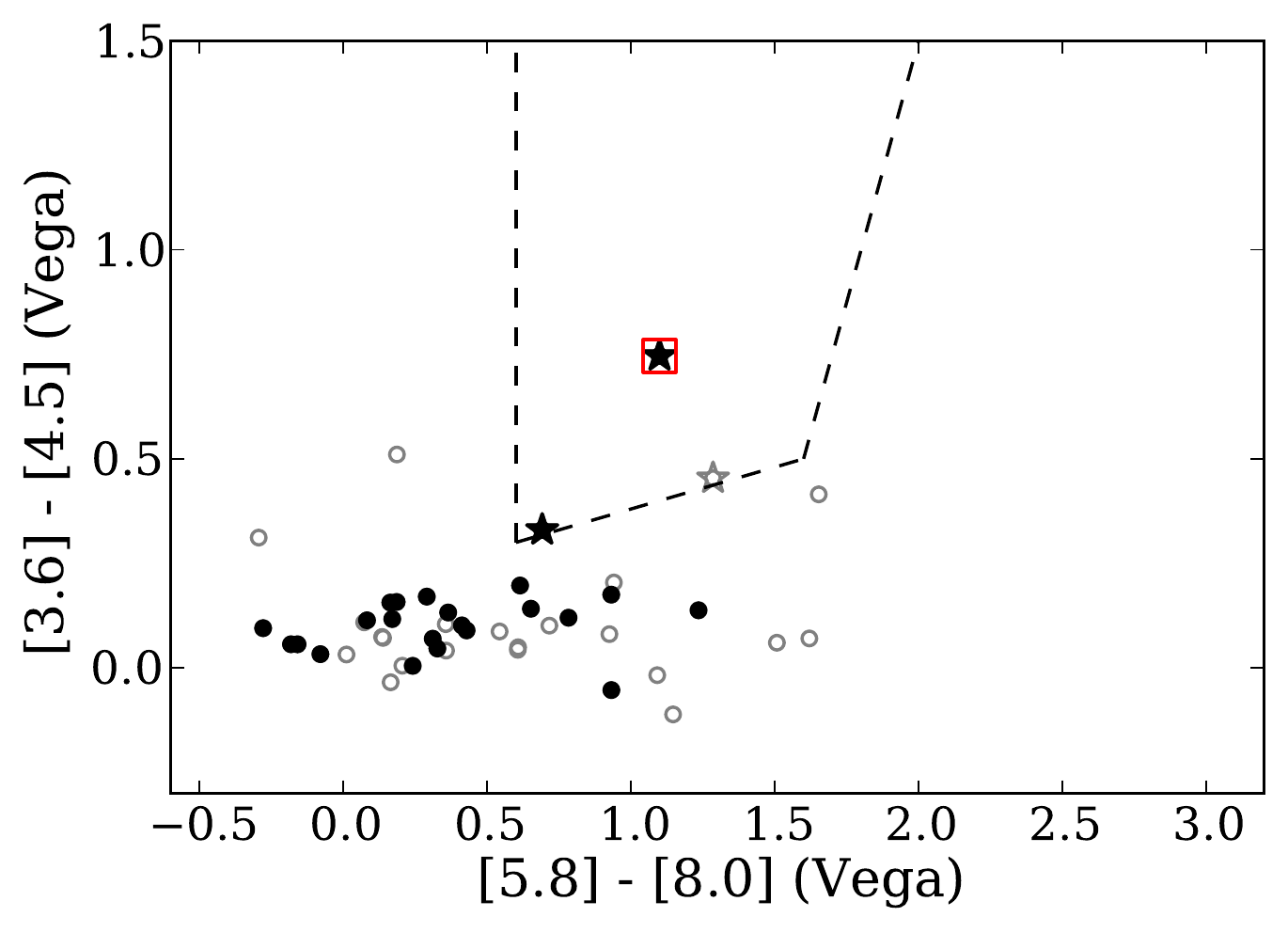}
\caption{AGN-SF distinction diagnostic based on IRAC \ciraca\ and \ciracb\ colours. Filled and open symbols represent FIR-detected and undetected sources. Stars are putative AGNs according to \cite{Stern2005}, while circles are SF galaxies. The red square marks the only FIR emitter within the AGN region according to the \cite{Donley2012} criterion. Note that magnitudes are reported in the Vega system.}\label{fig:agndiag}
\end{figure}

In addition to optical diagnostic methods, we can exploit the MIR AGN selection criterion proposed by \cite{Stern2005} which is based on the IRAC colours \ciraca\ and \ciracb.  This diagnostic is based on the fact that the MIR colours of SF galaxies are mainly caused by the different contributions from polycyclic aromatic hydrocarbons (PAHs), while those of AGNs are dominated by a power-law continuum up to $\lambda$\,$\sim$\,5\,\micron. As a result, the \ciraca\ colour is systematically higher while the \ciracb\ has a quite confined range. At most redshifts, the locus of AGNs is very clearly separated from that of SF galaxies. This diagnostic is more reliable for broad-line AGNs (BLAGN; $\sim$\,90\% accuracy) than for narrow-line AGNs (NLAGN; $\sim$\,40\%), but overall provides a reliable separation of AGNs from SF galaxies (and also for galactic stars), with 80\% completeness and less than 20\% contamination. The selection criterion proposed by \citet{Donley2012} is based on the same principles, but applies more stringent limits to avoid the contamination from SF galaxies when selecting the AGN population, at the cost of being highly incomplete at low AGN luminosities (less than 20\% at $\log L_x (erg s^{-1}) < 43$). 
We applied these diagnostics to our sample of cluster-candidate MIR-emitters as shown in Fig. \ref{fig:agndiag}. Of a total of 24 FIR emitters detected in the four IRAC channels, two objects fall within the Stern AGN region, and only one of them fulfils the Donley criterion. Given the very small numbers involved, the derived fractions of putative AGN should be taken with caution. They range from 10\% (considering the Stern criterion with 80\% completeness) to 20\% (applying the Donley limits assuming 20\% completeness). These fractions are similar to that obtained in \citet{PerezMartinez2012} and \citet{Lemaux2010}. 
On the other hand, remarkably, none of the individual X-ray point sources from \cite{Ulmer2009}  have a significant probability of being cluster members, meaning that the  MIR-selected AGNs do not show significant X-ray emission. We have highlighted the putative AGNs wherever appropriate, without removing them from the cluster SF sample because, on the one hand, their number is too small to have an impact on the statistical results, and on the other hand, it is very likely that a large portion of the FIR emission of these sources comes from SF activity.

	\section{Mapping the star-formation activity}
		\subsection{Local density} \label{subsec:ldens}
		
		To study the environmental dependence of the star formation activity, we use the nearest-neighbour density to characterise the environment \citep{Tanaka2005}. We defined a local density radius R$_{L}$ as the projected distance to the fifth nearest galaxy. Then, the local density was derived as $\Sigma_{5}$\,=\,6\,/\,$\pi$\,R$_{L}^{2}$, in Mpc$^{-2}$. In this calculation, we considered the full sample of 292 cluster member galaxies.
		
\begin{figure}
\centering
\includegraphics[width=\hsize]{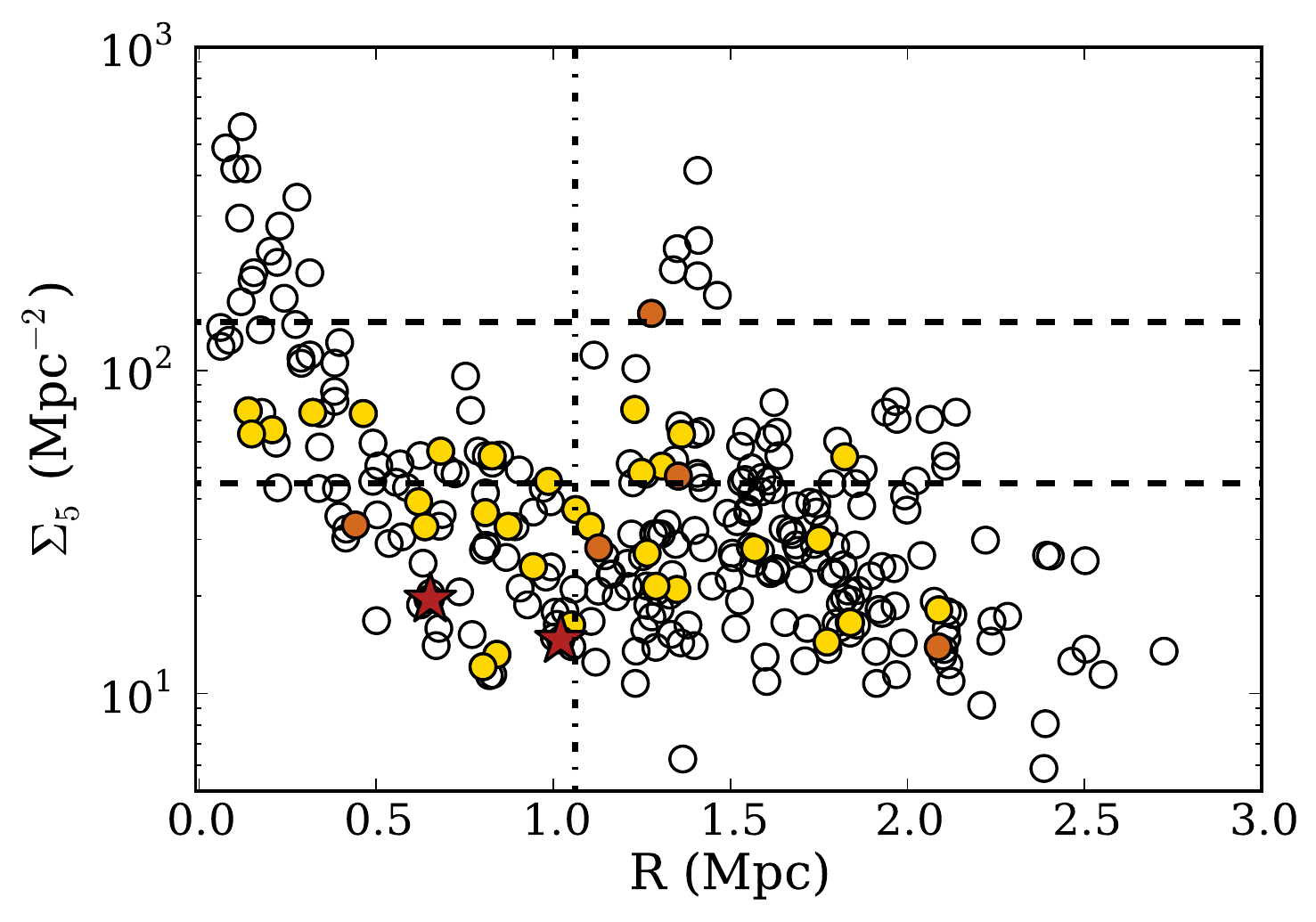}
\caption{Local density as a function of the cluster-centric radius. Open circles represent cluster members not detected in the FIR, yellow and orange filled circles represent MIPS-only and \herschel-detected star-forming galaxies, respectively. The vertical dash-dotted line at 1.05\,Mpc shows the virial radius. The horizontal dashed lines at log\,$\Sigma$\,=\,1.65 and 2.15 indicate the limits between the low-, intermediate-, and high-density environments. The red stars are the AGN candidates.}\label{fig:ldensr}
\end{figure}
		
		Following \citet[][hereafter KO08]{Koyama2008}, we considered three environmental bins. These authors defined the environmental regions considering the medium-density environment as a relative narrow range of local density where the optical colour distribution starts to change dramatically. The environments considered in KO08 are low-density (log$\Sigma_5$\,$<$\,1.65), intermediate-density (1.65\,$\leq$\,log$\Sigma_5$\,$<$\,2.15) and high-density (log$\Sigma_5$\,$\geq$\,2.15) regions. Since these definitions are arbitrary, we estimated its variation due to the modification of the density regions to study any property as a function of the local density, to ensure that the selected density regions do not condition our environmental dependence analysis. We changed the limits between regions within a Gaussian distribution centred on the KO08 limits definition, with a maximum variation of one sigma (fixed to 50\% the definition value). Thus, for each studied property our average resulting value in each density environment does not change with the region definition.
		
		 In Fig.\,\ref{fig:ldensr} we compare the local density with the clustercentric radius to determine the most appropriate parameter for studying the effect of the environment. The local density peaks not only at the centre but also around and beyond the virial radius show the high-density regions outside the cluster centre. Hence, we found that the local density is the most suitable parameter to account for the environment of RXJ1257.
		 

	   \subsection{Density map and structures}\label{subsec:structures}
\begin{figure*}
\centering
\includegraphics[width=0.49\hsize]{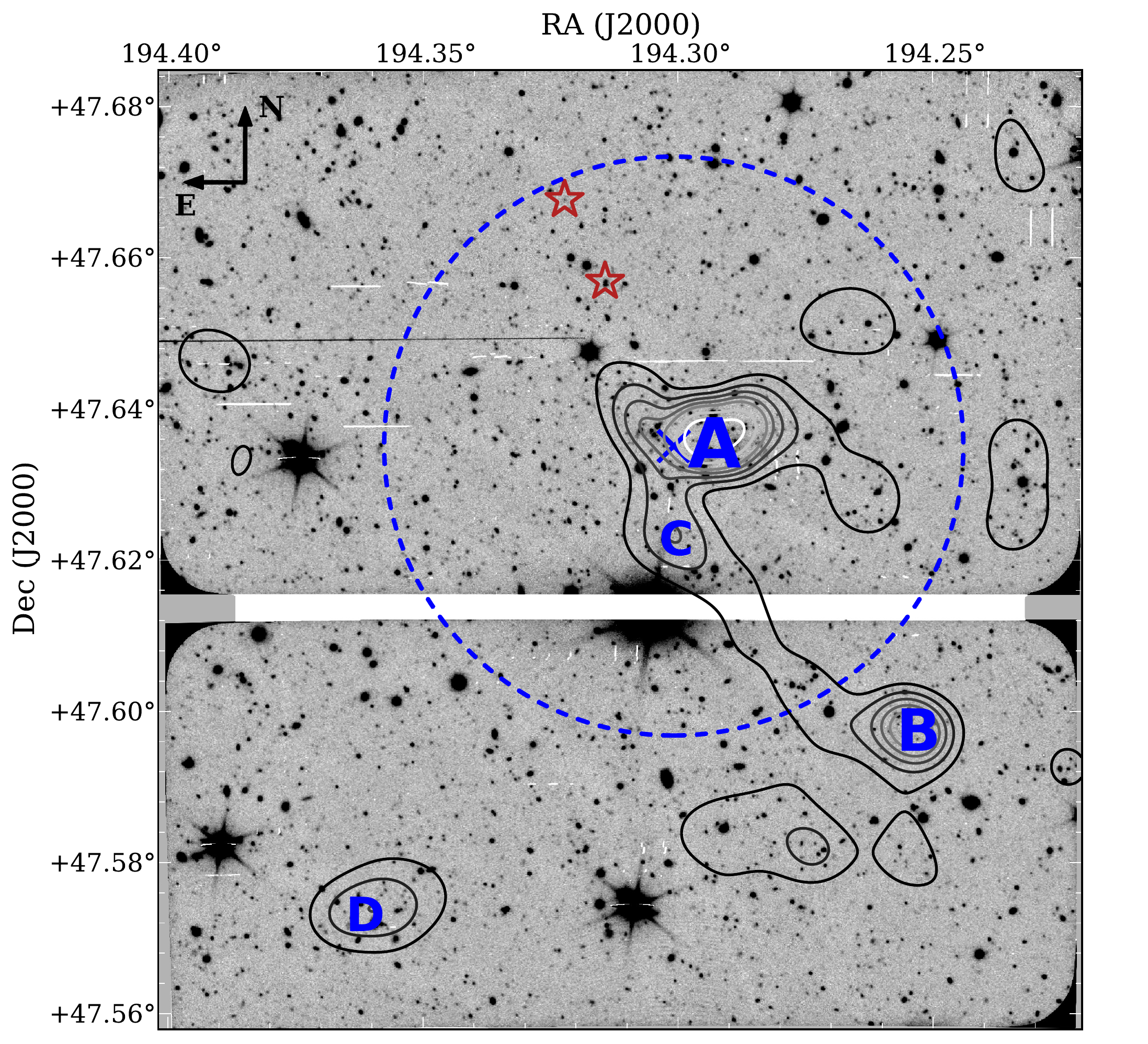}
\includegraphics[width=0.49\hsize]{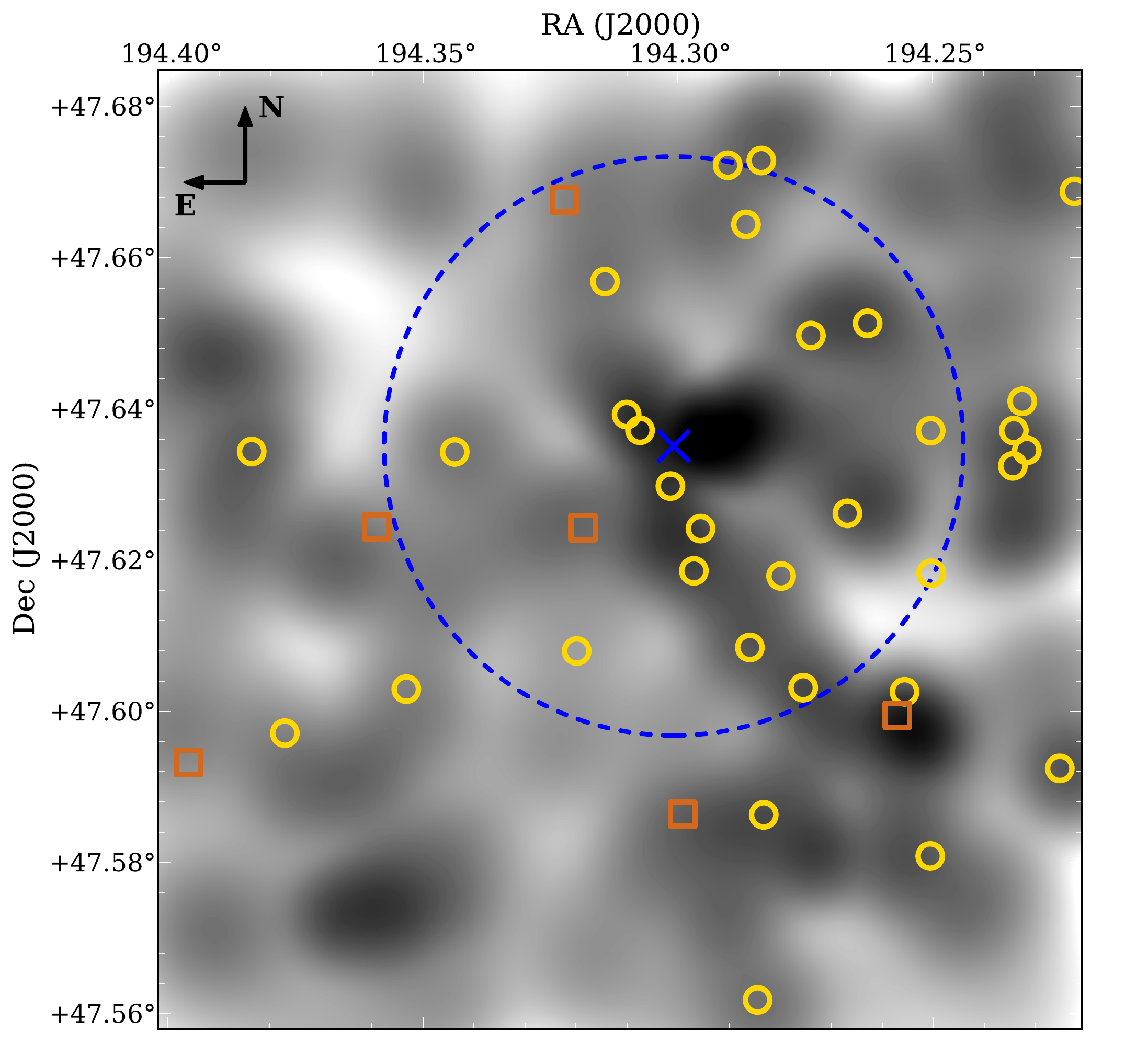}
\caption{(\textit{left}) Cluster structures on the r$'$-band OSIRIS/GTC image. The contours show the local density distribution of the cluster galaxies at  2, 3, 4, 5, 6, 8, and 10 times the background density noise, and the letters indicate the locus of the four structures detected above the 4$\sigma$ level with their size proportionally to the detection limit. The red stars show the AGN-candidate positions. (\textit{right}) Density map and spatial distribution of the FIR cluster members. Open yellow circles indicate the MIPS-only emitters and open orange squares represent \herschel-detected sources. The blue cross and dashed blue circle represent the centre and the virial radius of the cluster. }\label{fig:densdist}
\end{figure*}

	   To study the spatial density distribution of the cluster, we built a density map by applying an adaptive Gaussian kernel to smooth the cluster-member position distribution \citep{Pisani1996}. To properly describe the cluster spatial distribution, we defined the cluster structures as the regions with densities detected above the 4$\sigma$ level with respect to the background density fluctuation (see Table\,\ref{tab:structures} and letters in the left panel of Fig.\,\ref{fig:densdist}). These structures are found to remain the same throughout the Monte Carlo simulations performed in Sect.\,\ref{subsec:sim}. Moreover, these overdensities are kept invariable when the density maps are constructed from a subsample complete in stellar mass.
	   
	   In Fig.\,\ref{fig:densdist} the density map is shown in two ways: the left panel represents the density contours across the $r'$-band image and the right panel shows the FIR-emitter positions on the density map. Near the nominal cluster centre, an elongated structure (A) is almost aligned in the E-W direction. Within this structure, galaxies detected in MIPS 24\,$\mu$m-only appear in the NE part, suggesting that this elongation is a younger group in the process of assembling with the central region. Few overdensities, with FIR detected galaxies, are present around and beyond the virial radius in the EW direction and, especially,  in a filament-like structure in the SW direction. This filament-like structure, shown in the left panel of Fig.\,\ref{fig:densdist} with contours, includes the overdensities B and C, and the NE part of the structure A. The right panel of Fig.\,\ref{fig:densdist} shows that this filamentary structure is reproduced by the positions of the FIR emitters, making this structure the area of highest star formation. In contrast to region A, which shows the highest local density (detected at the 10$\sigma$ level), the two densest structures (B and C) were not reported by \cite{Ulmer2009}, probably because they were hidden by a mask in the focal plane. The FIR emitters in the densest regions and the filamentary distribution of the clusters structures suggest that RXJ1257 is a cluster in the process of formation, as has previously been pointed out by \cite{Ulmer2009}.
		
		\begin{table}[]
	    \caption{Cluster structures with densities detected above the 4$\sigma$ level. }\label{tab:structures}
		\begin{center}
		\begin{tabular}{l c c c c}
		\hline\hline
		\multirow{2}{*}{\textbf{ID}} & \textbf{R.A.} & \textbf{Dec} & \multirow{2}{*}{\textbf{Detection}} & \textbf{Pr. Distance} \\
		&  \textbf{(J2000)} & \textbf{(J2000)} & & \textbf{(Mpc)} \tablefootmark{a} \\
		\hline
		A & 12:57:10.30 & +47:38:08.232 & 10$\sigma$ & 0.149 \\
		B & 12:57:00.89 & +47:35:51.360 & 6$\sigma$ & 1.364 \\
		C & 12:57:12.19 & +47:37:22.656 & 4$\sigma$ & 0.338 \\
		D & 12:57:26.88 & +47:34:23.520 & 4$\sigma$ & 2.063 \\
		\hline
		\end{tabular}
		\tablefoot{\tablefoottext{a}{Measured relative to the nominal central position of the cluster (R.A.=12:57:12.2, Dec=+47:38:06.5)} }	
		\end{center}
		\end{table}		

	\subsection{Colour-density relation}

\begin{figure}
\centering
\includegraphics[width=\hsize]{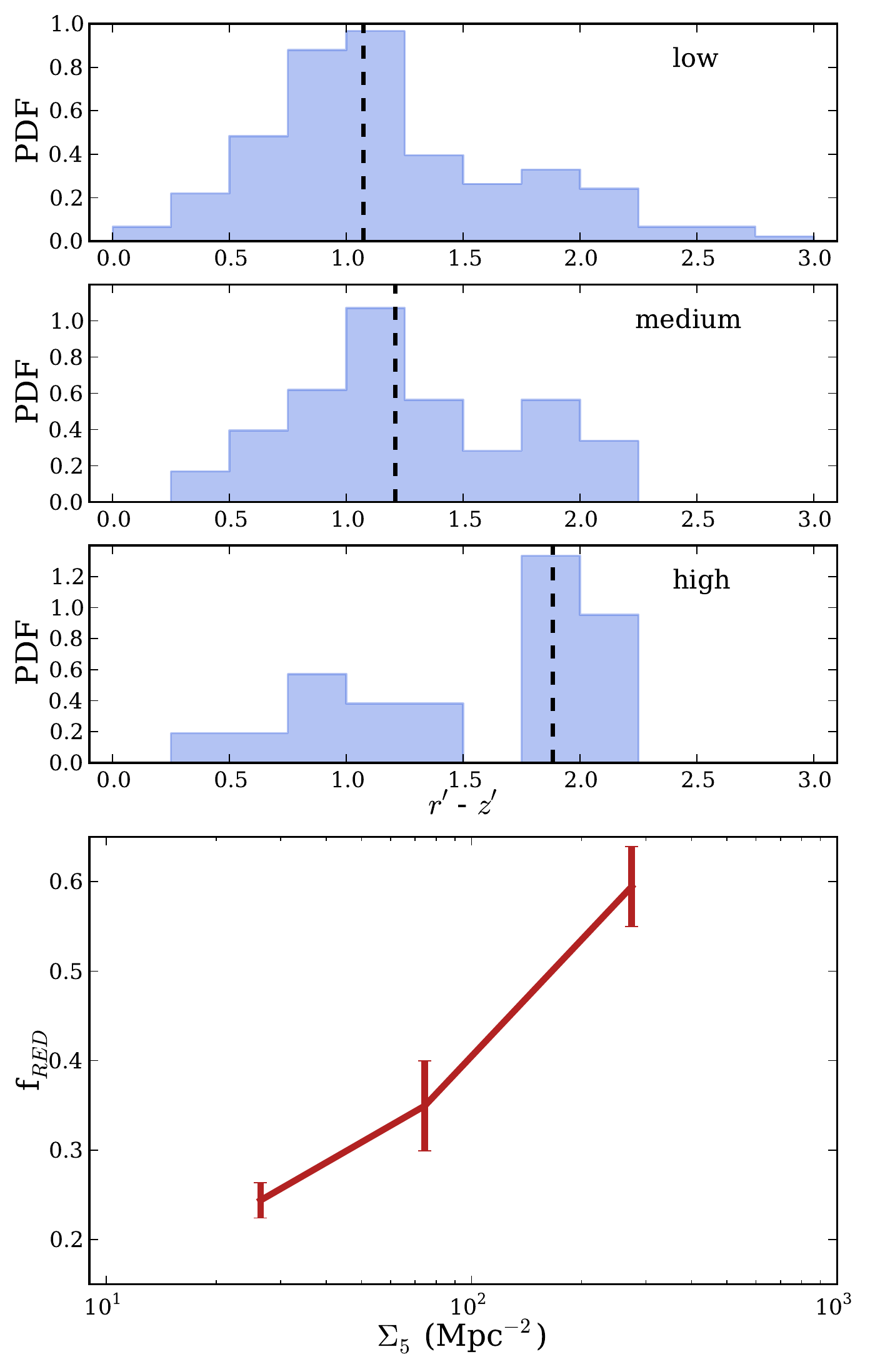}
\caption{(\textit{top}) Normalised $r'$\,-\,$z'$ colour distributions in each environment: low-, medium-, and high-density regions. The horizontal dashed lines indicate the median value of the $r'$\,-\,$z'$ colour in each panel. (\textit{bottom}) The fraction of red galaxies. Error bars show the standard deviation of the distribution of fractions of red galaxies in each density environment.}\label{fig:redseqColFrac}
\end{figure}

	Figure\,\ref{fig:redseqColFrac} shows the $r'$\,-\,$z'$ colour of individual cluster member galaxies as a function of the local density bins.  We observe an increment of the average colour with density, that is galaxies are redder towards denser environments, as can be seen from the black dashed line that represents the median in each environment, especially towards the high-density environment. These values were estimated as the median of the distribution of median colour values obtained by changing the limits between the originally defined environments, as explained in Sect.\,\ref{subsec:ldens}. This behaviour is observed for all cluster members (quiescent and SF galaxies) and for the FIR-emitter sample. This colour increment translates into an outstanding increase of the fraction of red galaxies relative to blue galaxies, as observed in the bottom panel of Fig.\,\ref{fig:redseqColFrac}: the fraction raises from $\sim$\,0.2 at low-density to $\sim$\,0.6 at high-density environments, reflecting the well-known morphology-density relation. The fraction of red galaxies was estimated as the median of the fraction distribution  (f$_{RED}$\,=\,n$_{RED}$/n$_{ALL}$) calculated by varying the limits between density regions. Error bars were computed with the standard deviation of the distribution of fractions.	
	
\begin{figure}
\centering
\includegraphics[width=0.9\hsize]{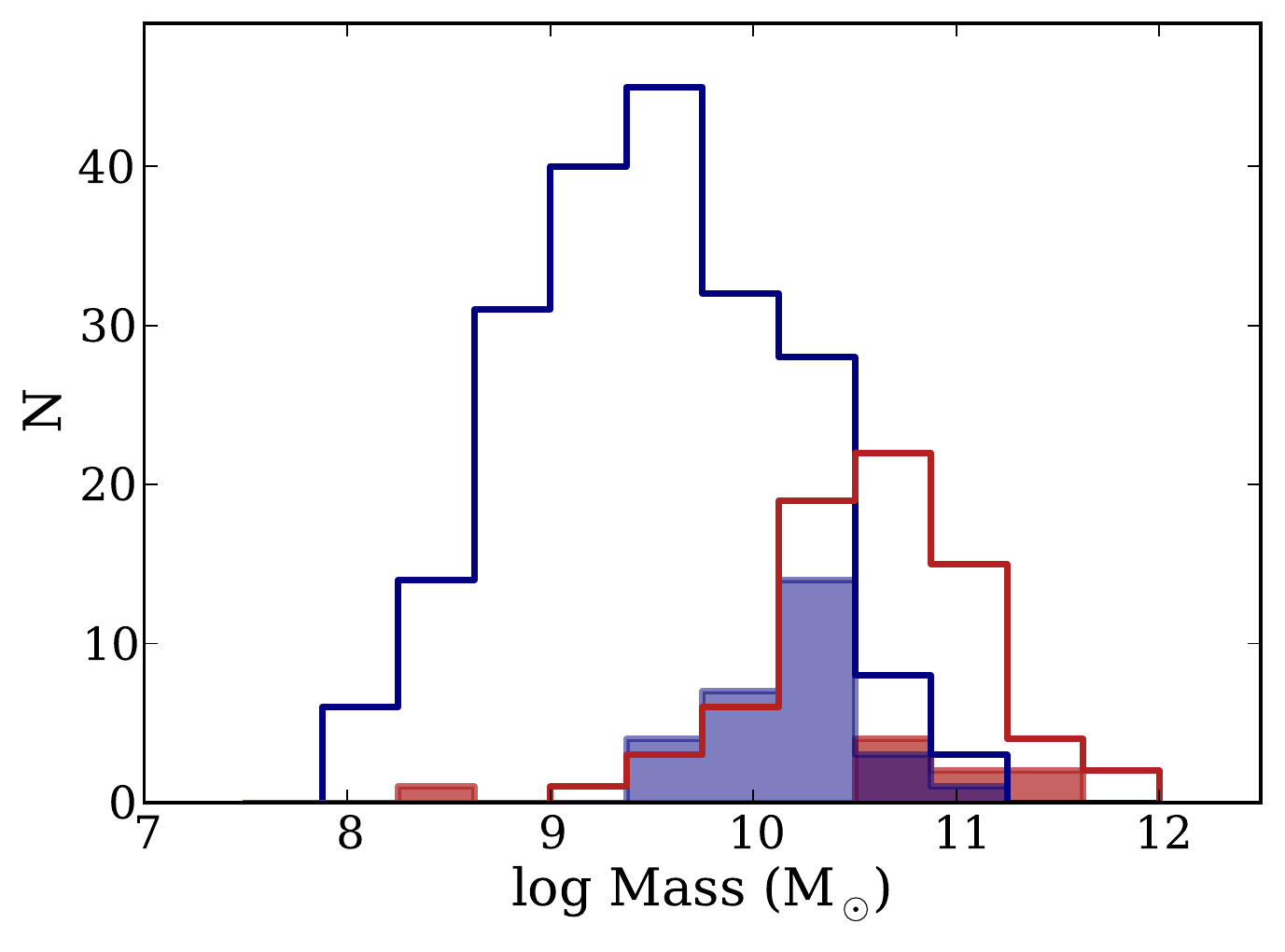}
\caption{Histograms of stellar mass distributions. Blue and red colours indicate distributions of galaxies belonging to the blue cloud and the red sequence, respectively, as defined in Sect.\,\ref{subsec:optcol}. Filled and open histograms show FIR-detected and undetected galaxies, respectively.}\label{fig:redbluemass}
\end{figure}

	In analysing the SFR dependence on the local density, we considered different categories, namely low- and high-mass galaxies and optically red and blue objects. These categories are not independent, as seen in Fig.\,\ref{fig:redbluemass}: red-sequence galaxies are generally found in the high-mass category, while blue cloud galaxies fall typically within the low-mass class. We decided to perform the analysis based on the high/low mass division and not red/blue galaxies to avoid biasing the sample with the $z'$ magnitude limit criterion used to define the red-sequence galaxies. The constraint imposed to define the red sequence additionally limit the sample and reduce the significance of the statistics. Although the observed trend indicates that massive galaxies tend to be red, the translation of the results obtained for high/low mass objects to red/blue ones can not be immediate and should be made with caution.  
	
	\subsection{Dependence of the star formation activity with environment}
	\label{subsec:sfdensity}
	
\begin{figure*}[ht]
\centering
\includegraphics[width=0.9\hsize]{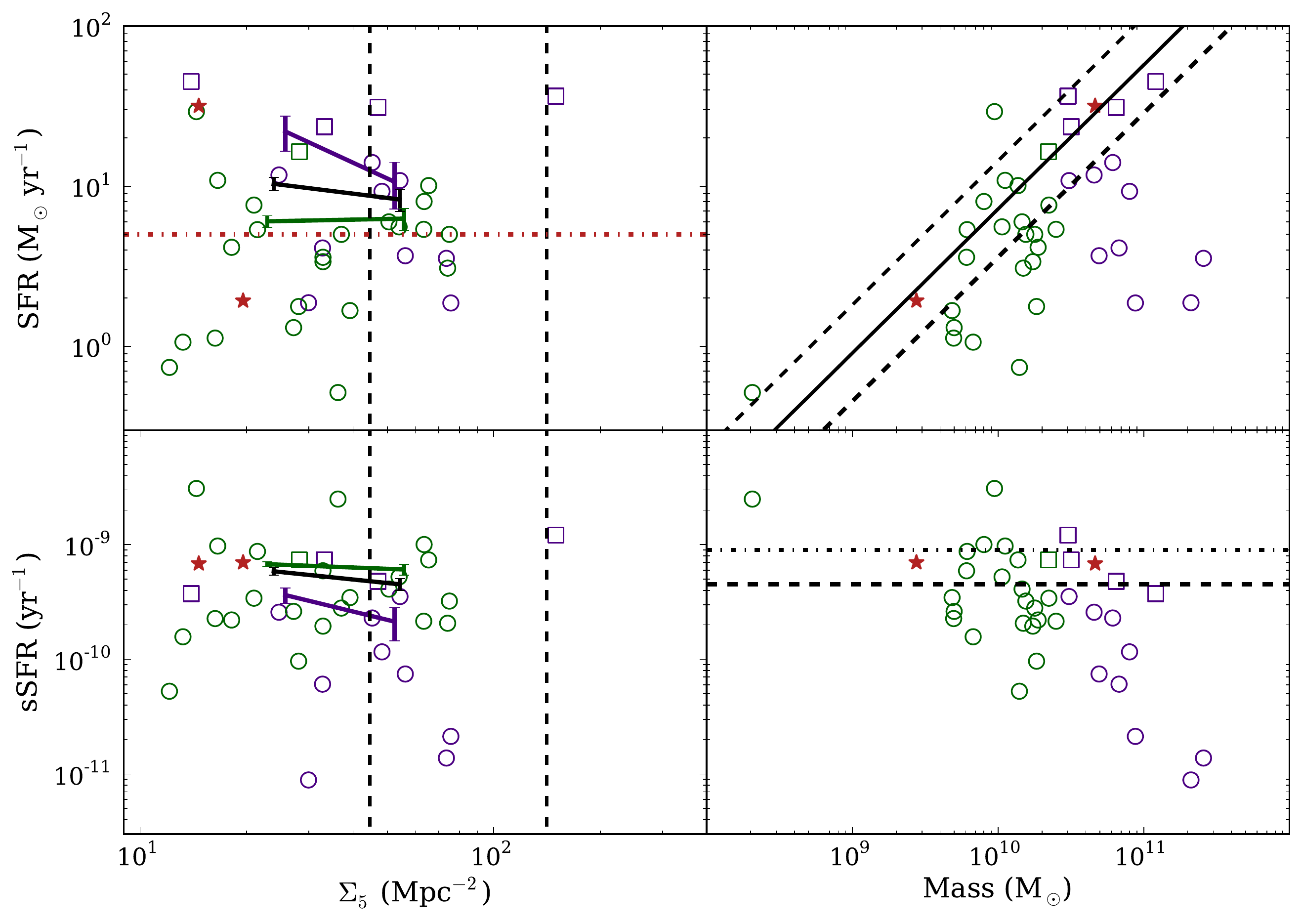}
\caption{SFR and sSFR of the FIR emitters as a function of the local density and the stellar mass. As in Fig.\,\ref{fig:ldensr}, open circles and open squares represent MIPS-only and \herschel-detected cluster members, respectively. Violet symbols show high-mass (M$_{\star}$\,$\geq$\,2.5\,$\times$\,10$^{10}$) galaxies, green symbols show low-mass (M$_{\star}$\,$<$\,2.5\,$\times$\,10$^{10}$) galaxies. In the left panels, we indicate the median values of the low-mass, the high-mass, and the full sample in each environment with green, violet, and black lines, respectively. The error bars are computed as the standard deviation of the median values distribution. The horizontal dot-dashed red line indicates the completeness limit of our FIR-emitter sample. The vertical dashed lines at log\,$\Sigma_5$\,=\,1.65 and 2.15 separate the low-, intermediate-, and high-density environments as defined in KO08. The average number of FIR-undetected galaxies is 190, 79, and 24 in the low-, intermediate-, and high-density bins, respectively. In the upper right pannel, the solid line indicates the SFR-M$_{\star}$ correlation from \citet{Elbaz2007}, and the dashed lines show the 68\% dispersion. In the lower right pannel, the dashed line corresponds to the relation of \citet{Elbaz2011} for the main-sequence galaxies at the redshift of the cluster, and the dashed-dotted line indicates the lower limit for the starburst region. The red stars in each panel show the AGN candidates.}\label{fig:sfrssfrall}
\end{figure*}

	In Fig.\ref{fig:sfrssfrall}, we plot the SFR and the sSFR of FIR emitters as a function of local density and stellar mass. The average SFR and sSFR were computed as the median of the distribution of median values obtained in each density bin by varying the boundaries of the regions, as explained in Sect.\,\ref{subsec:ldens}. Even with the strong variation of the region limits, we found that the median SFR and sSFR values are almost independent of the environment definition (as illustrated in Fig.\,\ref{fig:sfrssfrall}), the standard deviations of the median SFR and sSFR distributions are used as error bars for these average values. Due to the lack of FIR emitters in the high density environment we did not include the average value for this region.
	
	To separate the environment from the mass effect, we divided the sample of FIR emitters into low-mass (M$_{\star}$\,$<$\,2.5\,$\times$\,10$^{10}$) and high-mass (M$_{\star}$\,$\geq$\,2.5\,$\times$\,10$^{10}$) galaxies. This boundary is the characteristic mass at which the galaxy population changes from actively star-forming at lower stellar mass to typically quiescent at higher masses in the local Universe \citep{Kauffmann2003}. Despite the small number statistics of FIR emitters in each density bin, we see no significant environmental trends. Our FIR-emitter sample is highly biased towards high SFR galaxies, therefore the average value is not the best indicator for observing the environmental dependence of the star formation activity. We expect an environmental trend to be evident with a larger and more sensitive sample of SF galaxies.
	
	The analysis of the sSFR-density relation in two stellar mass bins implies a selection in sSFR due to the anticorrelation between the sSFR and the galaxy stellar mass (see right bottom panel of  Fig.\ref{fig:sfrssfrall}) \citep{Elbaz2007, Popesso2011}. The mean relation slightly anticorrelates with local density, following the observed trend with the stellar mass: low-mass SF galaxies are scarce in highest-density environments and dominate the high sSFR values in the low-density ones, while high-mass systems present a low sSFR and are the predominant SF galaxies observed in the highest-density environments.
	  	
\begin{figure}
\centering
\includegraphics[width=\hsize]{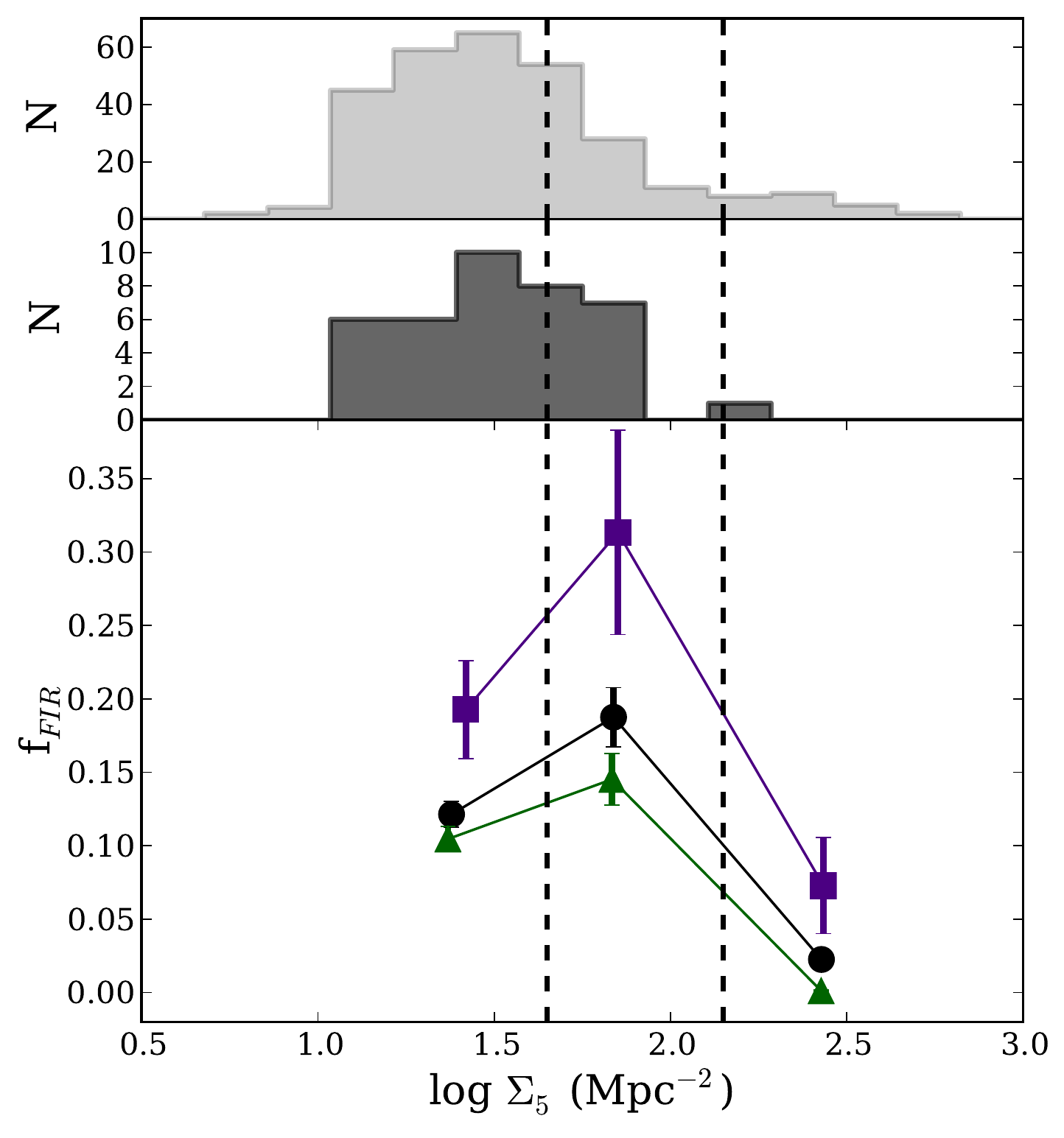}
\caption{Fractions of FIR emitters for low-mass galaxies (green triangles), high-mass galaxies (violet squares), and the full sample (black circles) as a function of the environment. Error bars are computed with Poisson statistics. The histograms in the upper panels represent the distribution of $\Sigma_5$ for the cluster members (\textit{top}) and for the FIR emitters (\textit{central}).}\label{fig:redseqFrac}
\end{figure}

	For each of the three environments, we derived the fraction of FIR emitters. For this calculation we used the full sample of 38 FIR emitters and the sample of 292 cluster member galaxies. Again, the fraction of FIR emitters was estimated as the median of the distribution of  fractions (f$_{FIR}$\,=\,n$_{FIR}$\,/\,n$_{ALL}$) calculated by varying the boundaries between density regions, as explained in Sect.\,\ref{subsec:ldens}. Error bars represent the Poisson statistics errors associated to the fractions. As can be seen in Fig. \ref{fig:redseqFrac}, we found that the fraction of FIR emitters increases at the medium density environment and rapidly decreases towards the high-density region. This behaviour is reproduced for both the low- and high-mass samples, showing no significant difference in the effect of the cluster environment as a function of the galaxy stellar mass. We found the same behaviour in the optically red and blue samples.
	
\begin{figure}
\centering
\includegraphics[width=0.9\hsize]{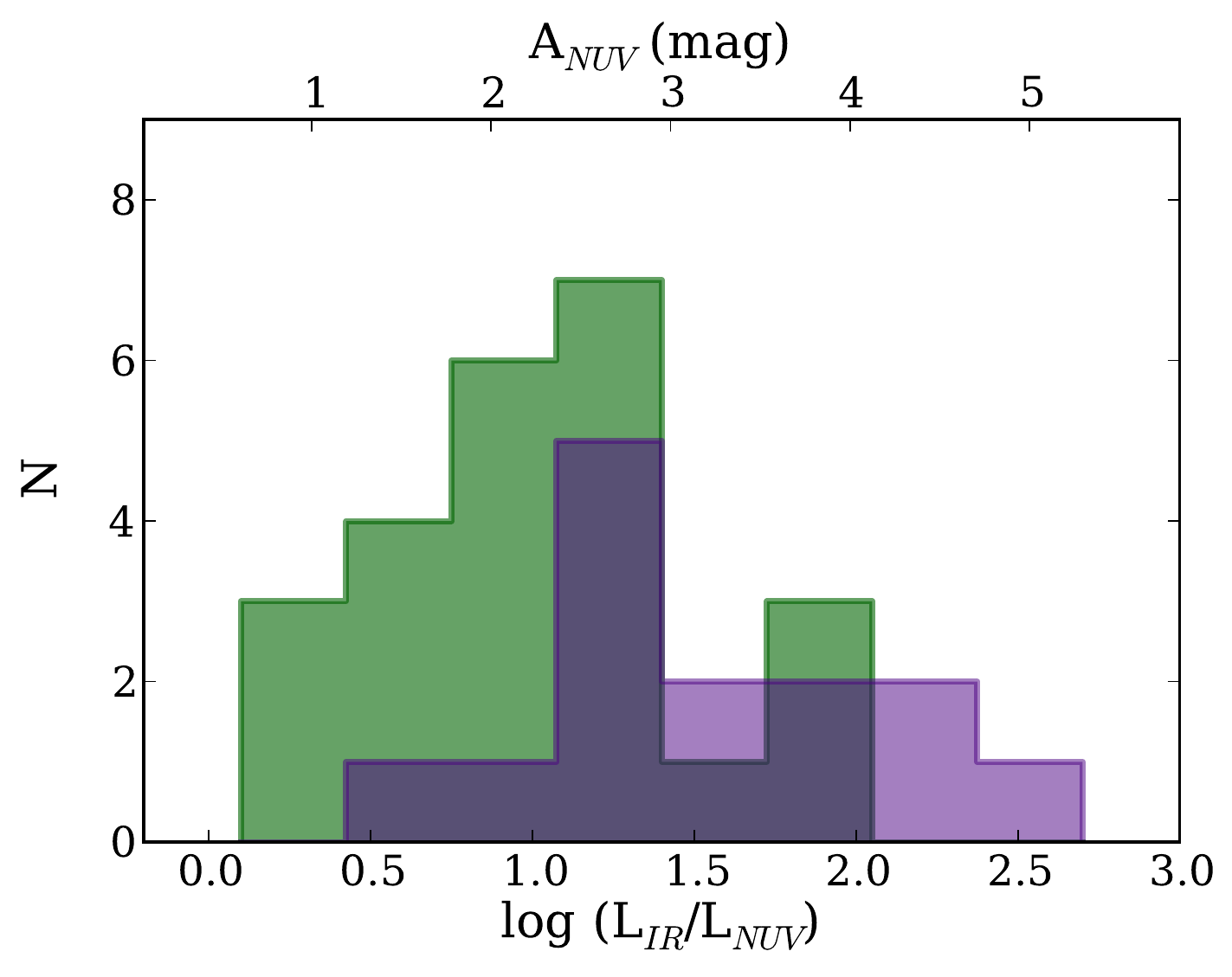}
\caption{The extinction distributions for the low (green) and high (violet) mass FIR emitters samples, estimated with the ratio between the total IR and UV luminosities. The top horizontal axis was calculated by using the calibration of \citet{Buat2005}.}\label{fig:redseqExtAge}
\end{figure}

	As explained in Sect.\,\ref{subsec:sedfitting}, extinction is one of the parameters we can obtain through the SED-fitting technique with \textit{LePhare}. However, the reduced wavelength range and the exponentially declining star formation histories used for the BC03 templates limit the reliability of these extinctions. A better-suited tracer of the dust attenuation is the ratio between the total IR and UV luminosities, which can be calibrated with \citet{Buat2005}. Figure\,\ref{fig:redseqExtAge} shows the dust attenuation estimated with this tracer for the low- and high-mass FIR-emitter samples. We observed that the distribution of high-mass galaxies is dustier. This difference was confirmed by a Kolmogorov-Smirnov (KS) test, since the statistics gives a probability below 4\% that the two distributions are drawn from the same parent population. The outcome of this analysis suggests that the high-mass SF galaxies are more dust-extincted than the low-mass sample. Equivalently, we have found the same result by comparing the optically red and blue SF galaxies, which indicates that the optical red colour of some FIR emitters may be due to dust attenuation.
	

	\section{Discussion}
	
	\subsection{Spatial distribution of FIR galaxies}
	The RXJ1257 cluster is formed by many projected substructures that are apparently in the process of merging. As detailed in Table \ref{tab:structures}, we have found four structures with a local density above 4$\sigma$ of the background level. The main structure (A) is very elongated in the W-NE direction: the central and western parts are characterised by the lack of FIR emitters, that is the SF in this region is completely quenched at our detection threshold; and the region towards the NE shows several FIR galaxies, which indicates that it is becoming the region with ongoing star formation.
	
	This scenario suggests an advanced infall-stage of the western region onto the high-density central structure, while the north-eastern region is experiencing a recent star-forming episode. This picture of a series of structures merging with an already evolved cluster does agree with the overabundance of red-sequence galaxies observed in and nearby the cluster central structure. The RXJ1257 cluster collapse has been reported by \citet{Ulmer2009} based on: \textit{(i)} a bimodal distribution of the X-ray emission in the galaxy population; \textit{(ii)} the number of MIR-detected spectroscopically confirmed cluster members; \textit{(iii)} the existence of substructures; and \textit{(iv)} the high kT compared to the L$_{X,bol}$-kT relation. We here suggested that the RXJ1257 cluster is currently in the process of formation. Nevertheless, while \citet{Ulmer2009} found MIPS cluster members associated with the central-western region, we did not spatially match those MIPS sources with spectroscopic cluster galaxies, but with photometric foreground or background objects and, moreover, we found MIPS-detected galaxies in the eastern group that match photometric cluster candidates.
	
	Another interesting feature of the projected spatial distribution is the overdensities alignment: B, C, and the north-eastern part of the A subgroups are aligned in the SW-NE direction. This spatial appearance looks more like a filament than a virialised symmetric structure, supporting the approach of studying the FIR emitters relative to their local density instead of their distance to the cluster centre. The filamentary structure formed by these overdensities is also reflected in the spatial distribution of the FIR emitters, as shown in the left panel of Fig.\,\ref{fig:densdist}, highlighting the medium density regions (filaments/group) as the preferred environment for the star formation activity.
	
	Previous similar studies in z\,$\sim$\,0.8 clusters have been performed by \citet{Marcillac2007} (RXJ0152 at z\,=\,0.83), \citet{Bai2007} (MS1054 at z\,=\,0.83), and KO08 (RXJ1716 at z\,=\,0.81). These works showed that SF cluster galaxies detected in the IR are distributed avoiding high-density regions and are detected especially in group/filament environments. Our work confirmed that FIR galaxies are located outside the cluster central region (A) and are preferentially distributed in intermediate densities.
	
		\subsection{Environmental dependence of the star formation activity}
    	As shown in Fig.\,\ref{fig:sfrssfrall},  neither the median SFR nor the median sSFR of FIR-detected SF galaxies show a strong dependence on the cluster environment, at least in the low- to intermediate-density media. Only when the sample is divided into low- and high-mass galaxies, a mild decrease of the median SFR of high mass objects is observed, although this is not very significant compared with the error bars. On the other hand, the SF is so suppressed for high-density regions that no statistical analysis is possible.
    	
        More interesting is the analysis of the dependence of the fraction of FIR-detected SF galaxies on the environment, as shown in Sect.\,\ref{subsec:sfdensity} and depicted in Fig.\,\ref{fig:redseqFrac}. A huge increase in the fraction  of FIR emitters is observed at intermediate densities while it falls to almost zero towards high-density environments, which shows that the cluster intermediate-density regions favour the star formation activity; this confirms previous results. At lower redshift, \citet{Biviano2011} found that the filament region in Abell 1763 hosts the highest fraction of IR emitters, but no environmental trend with their intrinsic properties was observed, that is, these authors obtained a similar sSFR vs M$_{\star}$ relation and SED galaxy classes for the different cluster environments. The RXJ1257 cluster shows a similar behaviour, although in this case the total SFR per number galaxy in each environment shows such a dispersion that we cannot claim that the intermediate density region contains the highest average SFR. The fact that we observed an environmental dependence of the fraction of FIR emitters, but not of their average SFR/sSFR, could be due to two main causes. On the one hand, the small number of objects available for tracing the transition region between the low- and intermediate-density environments. We roughly estimated the characteristic distance between these environments as $<$\,d\,$>$\,=\,0.2\,Mpc; considering the mean density of SF galaxies, we obtained that there are only approximately three sources to trace this transition region. Moreover, these sources are highly biased to high SFRs. On the other hand, as suggested by previous works \citep[e.g.][]{Biviano2011}, the quenching/enhancement of the star formation in clusters could be a fast process. For a typical galaxy velocity of 600\,km/s \citep{Ulmer2009}, the crossing time of the rough characteristic distance between the low- and intermediate-density environments is $\sim$\,0.3\,Gyr. Because we failed to detect any environmental dependence of the average SFR/sSFR, this time can be considered an upper limit for the duration of the physical mechanisms involved in the transformation of the galaxies in the transition region. Therefore, we need more sensitive data to assemble a detailed picture of the dependence of the SFR on the environment and to constrain the possible physical processes.

        The increase in the fraction of SF galaxies towards intermediate-density regions is found in both low- and high-mass galaxies, and an identical behaviour is observed in blue and red galaxies. The nature of the optically red SF galaxies has been discussed by KO08. The authors interpreted the red colours of some of their 15$\mu$m-emitting galaxies as due to dust reddening. We have investigated this possibility by comparing the distribution of extinction (as traced by the ratio L$_{IR}$/L$_{NUV}$) in blue and red SF objects (see the last part of Sect.\,\ref{subsec:sfdensity}). Since the only reliable estimate of the dust obscuration is derived through the total IR luminosity, we cannot compare the dust content of SF and quiescent galaxies. As a result of the comparison between red and blue SF galaxies, we found that it is very likely that the two distributions of dust attenuation are different. This finding supports the scenario that describes the optically red SF galaxies as dustier than the blue ones.

	\subsection{Global evolution}
\begin{figure}
\centering
\includegraphics[width=\hsize]{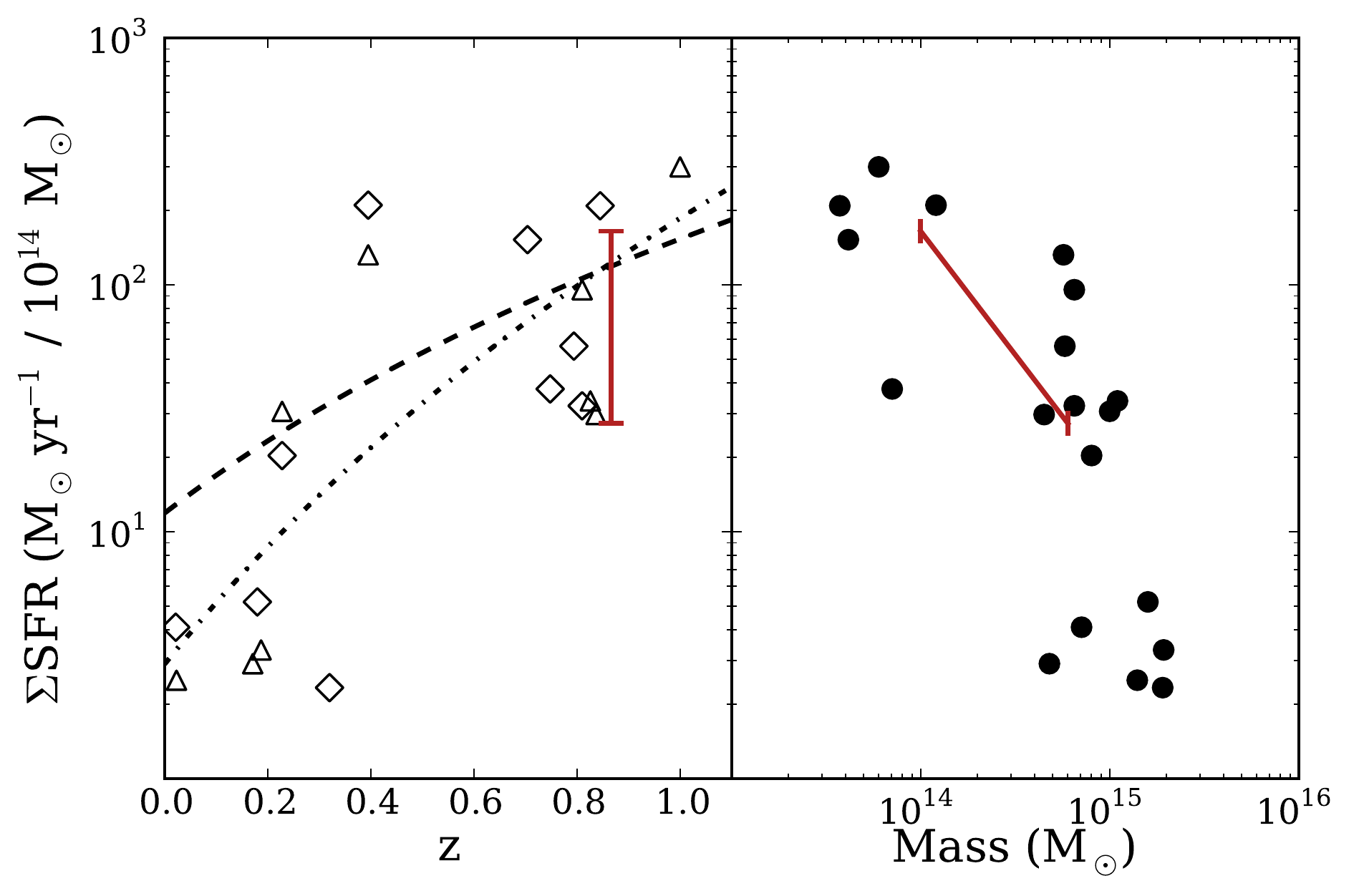}
\caption{Total cluster SFR per unit cluster mass as a function of redshift (\textit{left}) and as a function of cluster mass (\textit{right}). Open diamonds and triangles show the integrated total SFR estimated with H$\alpha$ and IR data, respectively, compiled by \citet{Koyama2010}. The red line indicates the cluster mass and, consequently, the$\Sigma$SFR/M$_{Cl}$ range of RXJ1257. The dashed line shows the best fit to $\Sigma$SFR/M$_{Cl}\sim(1+z)^\alpha$, and the dot-dashed line shows the best fit setting the power index to $\alpha$\,=\,6.}\label{fig:totalsfr}
\end{figure}

	We characterised the global star-forming properties of RXJ1257 in terms of the mass-normalized total SFR, $\Sigma$SFR/M$_{Cl}$, to compare them with previous studies in the literature. Following \citet{Koyama2010}, we estimated the $\Sigma$SFR using the cluster member galaxies within half the $R_{200}$ and divided it by $M_{200}$ to derive  $\Sigma$SFR/M$_{Cl}$. The radius and mass were calculated based on a velocity dispersion of 600\,km/s \citep{Ulmer2009}, and following equations \ref{eq:r200} and \ref{eq:m200},
	\begin{equation}
	R_{200}=2.47\dfrac{\sigma}{1000\,\rm km/s}\dfrac{1}{\sqrt{\Omega_{\Lambda 0}+\Omega_{m0}\left( 1+z\right)^{3}}}\,\rm Mpc
    \label{eq:r200}
    \end{equation}
	\begin{equation}
	M_{200}=1.71\times10^{15}\left( \dfrac{\sigma}{1000\,\rm km/s}\right)^{3}\dfrac{1}{\sqrt{\Omega_{\Lambda 0}+\Omega_{m0}\left( 1+z\right)^{3}}}\,\rm M_{\odot}.
	\label{eq:m200}
	\end{equation}
	We also considered the cluster mass calculated by \citet{Ulmer2009} with X-ray to estimate a range for the mass-normalized total SFR, obtaining $\Sigma$SFR/M$_{Cl}$\,=\,$[117,704]$\,M$_{\odot}$yr$^{-1}$\,/\,10$^{14}$\,M$_{\odot}$. In Fig.\,\ref{fig:totalsfr} we include RXJ1257 in the sample compiled by \citet{Koyama2010} (see references therein). We plot the cluster total SFR based on both H$\alpha$ intensity and MIR data.The mass-normalised total SFR increases towards higher redshift clusters and, although the scatter is large, this evolutionary trend can be fitted with $\propto(1+z)^6$.  When we fitted the power index as a free parameter, we obtained the best power index to be 4.4, still higher than the decline since $z\sim1$ found in field studies $\propto(1+z)^3$. Therefore, although a larger sample of clusters is needed, the cluster evolution supports a scenario in which the galaxy evolution in clusters is faster than that in the general field since $z\sim1$. We also confirm previous signs indicating that $\Sigma$SFR/M$_{Cl}$ is lower for systems of higher mass \citep{Finn2005,Bai2007,Koyama2010,Popesso2012}, as shown in the range in the right panel of Fig.\,\ref{fig:totalsfr}.

   \section{Summary and conclusions}
   We have performed a multi-wavelength optical and infrared study of RXJ1257.2+4738, a massive cluster at z=0.866. New OSIRIS/GTC and \herschel\ imaging data, along with ancillary data, have enabled us to perform SED-fitting and Monte Carlo simulations to build a catalogue of robust cluster members. This catalogue contains 292 cluster galaxies, 21 of which are spectroscopically confirmed cluster members and 271 are trustworthy photometric candidates (38 of them are FIR-detected). The identification of a large number of cluster galaxies has allowed us to investigate galaxy properties as a function of the cluster environment. Our main results are summarised below.
   \begin{enumerate}
   \item We calculated L$_{IR}$ luminosities for each of the 38 FIR-detected galaxies, obtaining a sample that is complete in IR luminosity down to L$_{IR}$\,$=$5.2\,$\times$\,10$^{10}$\,L$_{\odot}$, i.e. SFR\,$=$\,5\,M$_{\odot}$yr$^{-1}$. Twelve galaxies of this sample are in the LIRG range. Most of the FIR emitters fall into the main-sequence regime.
   \item We estimated that around 8\% of FIR emitters are AGN candidates, using a selection criterion based on the IRAC colours, and we confirm that none of them have X-ray emission. The resulting two candidates are located in the low-density regime, both are MIPS-detected and one has \herschel\ detected emission.
   \item We found four structures in the cluster density map with a local density above four times the background density fluctuation. The filament-like appearance of some overdensities and the spatial distribution of the FIR emitters within them suggest that RXJ1257 is still in the process of formation.
   \item We analysed the dependence of the star formation activity with the environment in three density regions (low-, intermediate-, and high-densities) and two mass samples (low- and high-mass galaxies). The average value of the SFR and the sSFR was shown to be roughly independent of the environment, although we should bear in mind that the data are highly biased to high SFRs. On the other hand, the fraction of star-forming galaxies was found to change with environment, showing the same behaviour for low- and high-mass samples. We observed that the fraction of low-mass SF galaxies increases from low- to intermediate-density environments and falls almost to zero  towards the densest regions. Both the spatial distribution and the fraction of FIR emitters lead us to consider the intermediate-density cluster environment as the appropriate region to observe the enhancement of star formation activity. The lack of an environmental dependence of the average SFR/sSFR of the FIR emitters against the dependence of their fraction suggests that the physical mechanism causing the variation of the star formation activity at intermediate densities could be a fast process.
   \item The analysis of the extinction distributions of the optically red and blue SF galaxies supports the assumption of the red SF galaxies as a dusty population. Nevertheless, to explore the origin of these optically red SF galaxies we need to separate the SED-fitting degeneration between extinction, age, and metallicity.
   \item The global star forming properties of RXJ1257 were characterised by estimating a range for the mass-normalised total SFR, $\Sigma$SFR/M$_{Cl}$\,=\,$[117,704]$\,M$_{\odot}$yr$^{-1}$\,/\,10$^{14}$\,M$_{\odot}$. The values within the obtained range are consistent with the scenario in which the evolution of cluster galaxies is faster than the evolution of field galaxies since z\,$\sim$\,1.
   \end{enumerate}
   
    A thorough study of the properties of the RXJ1257 cluster derived from optical emission lines (\oii, \oiii, \Hb), namely star formation down to $\sim$\,2\,M$_{\odot}$yr$^{-1}$, AGN activity and metallicity, along with a detailed morphological analysis will be presented in forthcoming papers.

    \begin{acknowledgements}
    We acknowledge the anonymous referee for helpful comments and suggestions that have greatly contributed to improve the manuscript.
    This research has been supported by the Spanish Ministry of Economy and Competitiveness (MINECO) under the grant AYA2011-29517-C03-01.
    We acknowledge support from the Faculty of the European Space Astronomy Centre (ESAC). 
    Based on observations made with the Gran Telescopio Canarias (GTC), instaled in the Spanish Observatorio del Roque de los Muchachos of the Instituto de Astrof\'isica de Canarias, in the island of La Palma.
    Based on observations made with the William Herschel Telescope operated on the island of La Palma by the Isaac Newton Group in the Spanish Observatorio del Roque de los Muchachos of the Instituto de Astrof\'isica de Canarias.
   PACS has been developed by a consortium of institutes led by MPE (Germany) and including UVIE (Austria); KUL, CSL, IMEC (Belgium); CEA, OAMP (France); MPIA (Germany); IFSI, OAP/AOT, OAA/CAISMI, LENS, SISSA (Italy); IAC (Spain). This development has been supported by the funding agencies BMVIT (Austria), ESA-PRODEX (Belgium), CEA/CNES (France), DLR (Germany), ASI (Italy) and CICYT/MICINN (Spain).
   SPIRE has been developed by a consortium of institutes led by Cardi Univ. (UK) and including: Univ. Lethbridge (Canada); NAOC (China); CEA, LAM (France); IFSI, Univ. Padua (Italy); IAC (Spain); Stockholm Observatory (Sweden); Imperial College London, RAL, UCL-MSSL, UKATC, Univ. Sussex (UK); and Caltech, JPL, NHSC, Univ. Colorado (USA). This development has been supported by national funding agencies: CSA (Canada); NAOC (China); CEA, CNES, CNRS (France); ASI (Italy); MCINN (Spain); SNSB (Sweden); STFC, UKSA (UK); and NASA (USA).
    \end{acknowledgements}

\bibliographystyle{aa} 
\bibliography{referencias} 

\onecolumn
\begin{landscape} 
\begin{center}

\begin{longtable}{l c c c c c c c c c c}
\caption{\label{tab:firdata} Properties of the cluster FIR emitters.} \\
\hline
\hline
\multirow{2}{*}{ID} & RA & Dec & \multirow{2}{*}{z} & \multirow{2}{*}{Flag}\tablefootmark{a}  & $\log$ Age & $\log$ M$_{\star}$ & $\log$ L$_{IR}$ & SFR & SSFR & \multirow{2}{*}{E(B-V)} \\
 & (J2000) & (J2000) & & & (yr) & (M$_\odot$) & (L$_\odot$) & (M$_\odot$yr$^{-1}$) & (yr$^{-1}$) & \\
\hline
\endfirsthead
\caption{continued.}\\
\hline
\hline
\multirow{2}{*}{ID} & RA & Dec & \multirow{2}{*}{z} & \multirow{2}{*}{Flag}\tablefootmark{a}   & $\log$ Age  & $\log$ M$_{\star}$ & $\log$ L$_{IR}$ & SFR & SSFR & \multirow{2}{*}{E(B-V)} \\
 & (J2000) & (J2000) & & & (yr) & (M$_\odot$) & (L$_\odot$) & (M$_\odot$yr$^{-1}$) & (yr$^{-1}$) & \\
\hline
\endhead
\hline
\endfoot 
  g-00020 & 194.2251 & 47.5925 & 0.86\,$\pm$\,0.14 & G1 & 8.7\,$\pm$\,0.8 & 10.0\,$\pm$\,0.3 & 11.05\,$\pm$\,0.08 & 10.9\,$\pm$\,1.1 & -9.0\,$\pm$\,0.6 & 0.2\\
  g-00123 & 194.2506 & 47.5809 & 0.95\,$\pm$\,0.10 & G1 & 8.7\,$\pm$\,0.8 & 10.0\,$\pm$\,0.3 & 10.76\,$\pm$\,0.01 & 5.6\,$\pm$\,0.1 & -9.3\,$\pm$\,0.5 & 0.1\\
  g-00153 & 194.2556 & 47.6026 & 0.85\,$\pm$\,0.00 & SP & 9.3\,$\pm$\,0.3 & 10.9\,$\pm$\,0.2 & 10.29\,$\pm$\,0.13 & 1.9\,$\pm$\,0.2 & -10.7\,$\pm$\,0.0 & 0.1\\
  g-00226 & 194.3960 & 47.5932 & 0.82\,$\pm$\,0.07 & G2 & 8.3\,$\pm$\,0.8 & 11.1\,$\pm$\,0.3 & 11.67\,$\pm$\,0.34 & 45.1\,$\pm$\,14.8 & -9.4\,$\pm$\,0.4 & 0.4\\
  g-00319 & 194.3771 & 47.5971 & 0.87\,$\pm$\,0.05 & G1 & 8.7\,$\pm$\,0.7 & 10.0\,$\pm$\,0.2 & 11.48\,$\pm$\,0.05 & 29.2\,$\pm$\,1.8 & -8.5\,$\pm$\,0.4 & 0.3 \\
  g-00427 & 194.2991 & 47.5864 & 0.87\,$\pm$\,0.07 & G1 & 8.7\,$\pm$\,0.8 & 10.8\,$\pm$\,0.3 & 11.51\,$\pm$\,0.14 & 31.1\,$\pm$\,5.6 & -9.3\,$\pm$\,0.6 & 0.4\\
  g-00499 & 194.2844 & 47.5618 & 0.91\,$\pm$\,0.12 & G1 & 8.9\,$\pm$\,0.8 & 10.3\,$\pm$\,0.3 & 10.64\,$\pm$\,0.01 & 4.2\,$\pm$\,0.0 & -9.7\,$\pm$\,0.5 & 0.3\\
  g-00505 & 194.2832 & 47.5863 & 0.81\,$\pm$\,0.10 & G1 & 8.5\,$\pm$\,0.8 & 10.4\,$\pm$\,0.3 & 10.75\,$\pm$\,0.10 & 5.4\,$\pm$\,0.7 & -9.7\,$\pm$\,0.4 & 0.4\\
  g-00535 & 194.2754 & 47.6032 & 0.81\,$\pm$\,0.11 & G1 & 9.0\,$\pm$\,0.5 & 10.8\,$\pm$\,0.2 & 11.17\,$\pm$\,0.08 & 14.0\,$\pm$\,1.2 & -9.6\,$\pm$\,0.2 & 0.4\\
  g-00588 & 194.3199 & 47.6080 & 0.90\,$\pm$\,0.11 & G1 & 8.4\,$\pm$\,0.8 & 9.8\,$\pm$\,0.3 & 10.04\,$\pm$\,0.01 & 1.1\,$\pm$\,0.0 & -9.8\,$\pm$\,0.3 & 0.3\\
  g-00665 & 194.3533 & 47.6030 & 0.92\,$\pm$\,0.14 & G2 & 8.5\,$\pm$\,0.8 & 10.3\,$\pm$\,0.3 & 10.90\,$\pm$\,0.01 & 7.6\,$\pm$\,0.1 & -9.5\,$\pm$\,0.3 & 0.4\\
  g-00795 & 194.2221 & 47.6688 & 0.88\,$\pm$\,0.09 & G1 & 9.7\,$\pm$\,0.3 & 11.3\,$\pm$\,0.2 & 10.29\,$\pm$\,0.01 & 1.9\,$\pm$\,0.0 & -11.1\,$\pm$\,0.4 & 0.0\\
  g-00850 & 194.2315 & 47.6345 & 0.89\,$\pm$\,0.08 & G1 & 8.4\,$\pm$\,0.7 & 10.\,$\pm$\,0.3 & 10.80\,$\pm$\,0.01 & 6.0\,$\pm$\,0.1 & -9.4\,$\pm$\,0.3 & 0.3\\
  g-00859 & 194.2324 & 47.6410 & 0.87\,$\pm$\,0.16 & G2 & 8.4\,$\pm$\,0.8 & 9.8\,$\pm$\,0.3 & 10.75\,$\pm$\,0.10 & 5.4\,$\pm$\,0.5 & -9.1\,$\pm$\,0.4 & 0.4\\
  g-00865 & 194.2343 & 47.6325 & 0.87\,$\pm$\,0.08 & G1 & 9.0\,$\pm$\,0.8 & 10.9\,$\pm$\,0.3 & 10.99\,$\pm$\,0.10 & 9.3\,$\pm$\,0.9 & -9.9\,$\pm$\,0.5 & 0.3\\
  g-00876 & 194.2341 & 47.6372 & 0.91\,$\pm$\,0.15 & G1 & 8.6\,$\pm$\,0.8 & 9.7\,$\pm$\,0.3 & 10.13\,$\pm$\,0.00 & 1.3\,$\pm$\,0.0 & -9.6\,$\pm$\,0.3 & 0.2\\
  g-00975 & 194.2503 & 47.6183 & 0.86\,$\pm$\,0.14 & G1 & 8.4\,$\pm$\,0.7 & 9.7\,$\pm$\,0.3 & 10.07\,$\pm$\,0.13 & 1.1\,$\pm$\,0.2 & -9.6\,$\pm$\,0.3 & 0.2\\
  g-00980 & 194.2504 & 47.6372 & 0.87\,$\pm$\,0.00 & SP & 9.0\,$\pm$\,0.4 & 10.7\,$\pm$\,0.1 & 11.09\,$\pm$\,0.08 & 11.7\,$\pm$\,1.0 & -9.6\,$\pm$\,0.2 & 0.3\\
  g-01044 & 194.2628 & 47.6514 & 0.82\,$\pm$\,0.11 & G2 & 8.5\,$\pm$\,0.8 & 10.5\,$\pm$\,0.3 & 11.05\,$\pm$\,0.08 & 10.8\,$\pm$\,0.9 & -9.5\,$\pm$\,0.4 & 0.2\\
  g-01245 & 194.3837 & 47.6344 & 0.91\,$\pm$\,0.10 & G1 & 9.3\,$\pm$\,0.8 & 10.3\,$\pm$\,0.3 & 10.27\,$\pm$\,0.00 & 1.8\,$\pm$\,0.0 & -10.0\,$\pm$\,0.6 & 0.3\\
  g-01402 & 194.3101 & 47.6393 & 0.85\,$\pm$\,0.09 & G1 & 8.3\,$\pm$\,0.8 & 10.1\,$\pm$\,0.3 & 11.02\,$\pm$\,0.08 & 10.1\,$\pm$\,0.9 & -9.1\,$\pm$\,0.3 & 0.5\\
  g-01419 & 194.3075 & 47.6372 & 0.92\,$\pm$\,0.12 & G2 & 8.6\,$\pm$\,0.8 & 10.2\,$\pm$\,0.3 & 10.72\,$\pm$\,0.02 & 5.0\,$\pm$\,0.1 & -9.5\,$\pm$\,0.4 & 0.2\\
  g-01463 & 194.3015 & 47.6298 & 0.89\,$\pm$\,0.15 & G1 & 8.3\,$\pm$\,0.7 & 9.9\,$\pm$\,0.2 & 10.92\,$\pm$\,0.01 & 8.0\,$\pm$\,0.1 & -9.0\,$\pm$\,0.2 & 0.2\\
  g-01495 & 194.2969 & 47.6186 & 0.86\,$\pm$\,0.00 & SP & 9.6\,$\pm$\,0.2 & 11.4\,$\pm$\,0.2 & 10.57\,$\pm$\,0.11 & 3.5\,$\pm$\,0.4 & -10.9\,$\pm$\,0.3 & 0.0\\
  g-01510 & 194.2956 & 47.6242 & 0.89\,$\pm$\,0.15 & G2 & 8.6\,$\pm$\,0.8 & 10.2\,$\pm$\,0.3 & 10.51\,$\pm$\,0.01 & 3.1\,$\pm$\,0.0 & -9.7\,$\pm$\,0.4 & 0.4\\
  g-01555 & 194.2902 & 47.6723 & 0.91\,$\pm$\,0.12 & G1 & 8.3\,$\pm$\,0.8 & 10.3\,$\pm$\,0.3 & 10.72\,$\pm$\,0.02 & 5.0\,$\pm$\,0.1 & -9.6\,$\pm$\,0.3 & 0.5\\
  g-01584 & 194.2866 & 47.6645 & 0.93\,$\pm$\,0.11 & G2 & 7.5\,$\pm$\,0.8 & 9.8\,$\pm$\,0.4 & 10.57\,$\pm$\,0.01 & 3.6\,$\pm$\,0.0 & -9.2\,$\pm$\,0.2 & 0.6\\
  g-01600 & 194.2836 & 47.6729 & 0.89\,$\pm$\,0.15 & G1 & 8.5\,$\pm$\,0.7 & 10.2\,$\pm$\,0.3 & 10.55\,$\pm$\,0.01 & 3.4\,$\pm$\,0.0 & -9.7\,$\pm$\,0.3 & 0.3\\
  g-01654 & 194.2739 & 47.6498 & 0.83\,$\pm$\,0.07 & G1 & 9.0\,$\pm$\,0.3 & 10.8\,$\pm$\,0.2 & 10.63\,$\pm$\,0.11 & 4.1\,$\pm$\,0.6 & -10.2\,$\pm$\,0.1 & 0.1\\
  g-01702 & 194.2668 & 47.6262 & 0.86\,$\pm$\,0.00 & SP & 9.0\,$\pm$\,0.7 & 10.7\,$\pm$\,0.3 & 10.58\,$\pm$\,0.11 & 3.7\,$\pm$\,0.4 & -10.1\,$\pm$\,0.4 & 0.4\\
  g-01770 & 194.3591 & 47.6244 & 0.86\,$\pm$\,0.12 & G1 & 8.3\,$\pm$\,0.5 & 10.3\,$\pm$\,0.2 & 11.23\,$\pm$\,0.33 & 16.4\,$\pm$\,9.0 & -9.1\,$\pm$\,0.5 & 0.5\\
  g-01773 & 194.3186 & 47.6243 & 0.83\,$\pm$\,0.08 & G1 & 8.5\,$\pm$\,0.8 & 10.5\,$\pm$\,0.3 & 11.39\,$\pm$\,0.31 & 23.5\,$\pm$\,8.4 & -9.1\,$\pm$\,0.5 & 0.5\\
  g-01858 & 194.3438 & 47.6344 & 0.86\,$\pm$\,0.00 & SP & 8.8\,$\pm$\,0.8 & 10.1\,$\pm$\,0.5 & 9.89\,$\pm$\,0.13 & 0.7\,$\pm$\,0.1 & -10.3\,$\pm$\,0.7 & 0.3\\
  g-01883 & 194.3223 & 47.6677 & 0.98\,$\pm$\,0.08 & G1 & 8.8\,$\pm$\,0.8 & 10.7\,$\pm$\,0.3 & 11.52\,$\pm$\,0.27 & 31.7\,$\pm$\,7.6 & -9.2\,$\pm$\,0.4 & 0.6\\
  g-00160 & 194.2570 & 47.5995 & 0.85\,$\pm$\,0.00 & SP & 8.5\,$\pm$\,0.8 & 10.5\,$\pm$\,0.3 & 11.58\,$\pm$\,0.37 & 36.5\,$\pm$\,9.4 & -8.9\,$\pm$\,0.4 & 0.4\\
  g-01391 & 194.3143 & 47.6569 & 0.86\,$\pm$\,0.00 & SP & 7.0\,$\pm$\,0.8 & 9.4\,$\pm$\,0.3 & 10.30\,$\pm$\,0.13 & 1.9\,$\pm$\,0.3 & -9.2\,$\pm$\,0.0 & 0.6\\
  g-01630 & 194.2798 & 47.6179 & 0.87\,$\pm$\,0.00 & SP & 8.7\,$\pm$\,0.7 & 9.7\,$\pm$\,0.3 & 10.24\,$\pm$\,0.13 & 1.7\,$\pm$\,0.2 & -9.5\,$\pm$\,0.4 & 0.2\\
\end{longtable}    
\tablefoot{\tablefoottext{a}{Redshift flag: spectroscopic redshift (SP), photometric redshift fitted with a Gaussian function (G1), and photometric redshift fitted with a double Gaussian function (G2). See Sect.\,\ref{sec:clusterdet} for more details about the redshift estimation.} }	
\end{center}
\end{landscape} 

\end{document}